\documentclass[hyper, letterpaper,11pt, notoc]{JHEP3}
\usepackage{graphicx}
\usepackage{amssymb}
\usepackage{amsmath}
\usepackage{latexsym}
\usepackage[all]{xy}

\newcommand{\be}{\begin{equation}}
\newcommand{\beq}[1]{\begin{equation}\label{#1}}
\newcommand{\eq}[1]{eq.~(\ref{#1})}
\newcommand{\eqs}[2]{eqs.\ (\ref{#1}) and (\ref{#2})}
\newcommand{\ee}{\end{equation}}
\newcommand{\eeq}{\end{equation}}
\newcommand{\bea}{\begin{eqnarray}}
\newcommand{\eea}{\end{eqnarray}}
\newcommand{\comment}[1]{}
\newcommand{\TeV}{~\mathrm{TeV}}
\newcommand{\GeV}{~\mathrm{GeV}}

\newcommand{\lsim}{\!\mathrel{\hbox{\rlap{\lower.55ex \hbox{$\sim$}} \kern-.34em \raise.4ex \hbox{$<$}}}}
\newcommand{\gsim}{\!\mathrel{\hbox{\rlap{\lower.55ex \hbox{$\sim$}} \kern-.34em \raise.4ex \hbox{$>$}}}}

\DeclareMathOperator{\diag}{diag}
\DeclareMathOperator{\tr}{tr}

\newcommand{\doublerightxyarrow}{\ar@{-}[rr] |-{\SelectTips{eu}{}\object@{>}}}
\newcommand{\doublerightdownxyarrow}{\ar@{-}[rdr] |-{\SelectTips{eu}{}\object@{>}}}
\newcommand{\doublerightupxyarrow}{\ar@{-}[rur] |-{\SelectTips{eu}{}\object@{>}}}

\newcommand{\rightxyarrow}{\ar@{-}[r] |-{\SelectTips{eu}{}\object@{>}}}
\newcommand{\rightdownxyarrow}{\ar@{-}[rd] |-{\SelectTips{eu}{}\object@{>}}}
\newcommand{\downxyarrow}{\ar@{-}[d] |-{\SelectTips{eu}{}\object@{>}}}
\newcommand{\rightupxyarrow}{\ar@{-}[ru] |-{\SelectTips{eu}{}\object@{>}}}
\newcommand{\upxyarrow}{\ar@{-}[u] |-{\SelectTips{eu}{}\object@{>}}}
\newcommand{\leftupxyarrow}{\ar@{-}[lu] |-{\SelectTips{eu}{}\object@{>}}}
\newcommand{\leftdownxyarrow}{\ar@{-}[ld] |-{\SelectTips{eu}{}\object@{>}}}
\newcommand{\leftxyarrow}{\ar@{-}[l] |-{\SelectTips{eu}{}\object@{>}}}

\newcommand{\longdashedrightxyarrow}{\ar@{--}[rr] |-{\SelectTips{eu}{}\object@{>}}}
\newcommand{\longdashedrightdownxyarrow}{\ar@{--}[rdrd] |-{\SelectTips{eu}{}\object@{>}}}
\newcommand{\longdasheddownxyarrow}{\ar@{--}[dd] |-{\SelectTips{eu}{}\object@{>}}}
\newcommand{\longdashedrightupxyarrow}{\ar@{--}[ruru] |-{\SelectTips{eu}{}\object@{>}}}
\newcommand{\longdashedupxyarrow}{\ar@{--}[uu] |-{\SelectTips{eu}{}\object@{>}}}
\newcommand{\longdashedleftupxyarrow}{\ar@{--}[lulu] |-{\SelectTips{eu}{}\object@{>}}}
\newcommand{\longdashedleftdownxyarrow}{\ar@{--}[ldld] |-{\SelectTips{eu}{}\object@{>}}}
\newcommand{\longdashedleftxyarrow}{\ar@{--}[ll] |-{\SelectTips{eu}{}\object@{>}}}

\newcommand{\altleftxyarrow}{\ar@{-}[r] |-{\SelectTips{eu}{}\object@{<}}}

\newcommand\openone{\leavevmode\hbox{\small1\normalsize\kern-.33em1}}

\preprint{HUTP-05/A0008}

\title{Little Technicolor}

\author{Jesse Thaler \\ Jefferson Physical Laboratory, Harvard University, Cambridge, MA 02138 \\ E-mail:  \email{jthaler@jthaler.net}}

\abstract{Inspired by the AdS/CFT correspondence, we show that any $G/H$ symmetry breaking pattern can be described by a simple two-site moose diagram.  This construction trivially reproduces the CCWZ prescription in the context of Hidden Local Symmetry.  We interpret this moose in a novel way to show that many little Higgs theories can emerge from ordinary chiral symmetry breaking in scaled-up QCD.  We apply this reasoning to the simple group little Higgs to see that the same low energy degrees of freedom can arise from a variety of UV complete theories.  We also show how models of holographic composite Higgs bosons can turn into brane-localized little technicolor theories by ``integrating in'' the IR brane.}

\begin{document}

\section{Introduction}

Whether or not strong dynamics ultimately explains the hierarchy problem, theories that exhibit confinement are undoubtedly the most elegant way to generate large hierarchies of scales \cite{Gross:1973id,Politzer:1973fx}.  In the context of the standard model, there is the unfortunate reality that the electroweak scale ($246 \GeV$), the precision electroweak scale ($\sim 10 \TeV$), and the flavor scale ($\sim 1000 \TeV$) are well separated, and no known strong dynamics can elegantly explain \emph{three} hierarchies.  In this light, little Higgs theories \cite{Arkani-Hamed:2001nc,Arkani-Hamed:2002pa,Arkani-Hamed:2002qy,Arkani-Hamed:2002qx,Gregoire:2002ra,Low:2002ws,Kaplan:2003uc,Skiba:2003yf} are a reasonable compromise to the hierarchy problem:  strong dynamics separate the precision electroweak scale from the Planck scale, collective breaking explains the ``little hierarchy'' \cite{Cheng:2003ju,Cheng:2004yc} between the confinement scale and the electroweak scale, and the flavor problem is presumably addressed by some sort of GIM mechanism \cite{Glashow:1970gm}.

For all the phenomenological successes of little Higgs theories, their structure can seem a bit artificial from the high energy perspective.  While the $SU(5)/SO(5)$ littlest Higgs \cite{Arkani-Hamed:2002qy} could arise from strong $SO(N)$ dynamics \cite{Katz:2003sn}, the $\left(SU(3)/SU(2)\right)^2$ simple group little Higgs \cite{Kaplan:2003uc} has no obvious QCD-like UV completion, because it is hard (though not impossible) to imagine that an ordinary confining theory would break an $SU(N)$ flavor symmetry to $SU(N-k)$.  In this light, technicolor theories \cite{Weinberg:1975gm,Susskind:1978ms} seem a lot more realistic (if not phenomenologically viable) in that they are simply scaled-up versions of ordinary $SU(N_f)_L \times SU(N_f)_R \rightarrow SU(N_f)_D$ chiral symmetry breaking in QCD.  But before we dismiss the little Higgs theories as interesting but artificial constructions, we will actually show that  \emph{any} $(SU(N)/H)^n$ little Higgs theory can arise from $n$ copies of QCD with $N$ flavors!  In other words, the little Higgs could literally be a pion of QCD.

We will show that an $SU(N)/H$ little Higgs theory where an $F$ subgroup of $SU(N)$ is gauged can be described by the moose diagram:
\beq{intromoose}
\begin{tabular}{c}
\xymatrix@R=.4pc@C=1.4pc{\mathrm{Global:} & SU(N)_L && SU(N)_R \\
& *=<20pt>[o][F]{} \rightxyarrow^{\mbox{\raisebox{1.5ex}{$\psi$}}} & *=<12pt>[o][F=]{} \rightxyarrow^{\mbox{\raisebox{1.5ex}{$\psi^c$}}} & *=<20pt>[o][F]{} \\ \mathrm{Gauged:} &F &SU(N_c)&H}
\end{tabular}
\ee
This is just QCD with $N$ flavors where some of the flavor symmetries have been gauged.   As we will see, this ``little technicolor'' moose emerges quite naturally from deconstructing the AdS dual of some quasi-CFT.  However, even without insight from AdS/CFT, we could study \eq{intromoose} in its own right as a novel UV completion of little Higgs theories.

The main results of this paper are contained in section \ref{sec:intro}:  we present the AdS/CFT inspiration for little technicolor and show how it connects to the known formalisms of CCWZ \cite{Coleman:1969sm,Callan:1969sn} and Hidden Local Symmetry (HLS) \cite{Bando:1987br}; we then reinterpret the HLS construction in a novel way to arrive at the little technicolor moose.  In section \ref{sec:four}, we apply the little technicolor construction to the $\left(SU(3)/SU(2)\right)^2$ simple group little Higgs to show how the same low energy degrees of freedom can arise from four very different theories:  a straightforward application of AdS/CFT, two copies of QCD, one copy of QCD in the vector limit \cite{Georgi:1989xy}, and a known AdS$_5$ construction \cite{Contino:2003ve} that superficially does not look like a little Higgs theory (but really is).  We return briefly to AdS space in section \ref{sec:holo} to show how ``integrating in'' the IR brane can turn holographic composite Higgs models into brane-localized little Higgs theories.  We comment on vacuum alignment issues in section \ref{sec:fvsh}, and we conclude with some outstanding questions about more general little Higgs theories and speculations on the Wess-Zumino-Witten term \cite{Wess:1971yu,Witten:1983tw}.

\section{From AdS/CFT to QCD via CCWZ and HLS}
\label{sec:intro}

The starting point for our analysis is the AdS/CFT correspondence \cite{Maldacena:1997re,Gubser:1998bc,Witten:1998qj} and its phenomenological interpretation \cite{Arkani-Hamed:2000ds,Rattazzi:2000hs}.   There is a straightforward way to construct the AdS dual of a CFT that yields a $G/H$ nonlinear sigma model at low energies and where a subgroup $F \subset G$ is gauged:  simply consider a slice of AdS$_5$ \cite{Randall:1999ee} with bulk $G$ gauge bosons where the gauge symmetry is reduced to $F$ on the UV brane and $H$ on the IR brane \cite{Contino:2003ve}:
\medskip
\beq{duality}
\begin{tabular}{c}
\includegraphics[scale=0.55]{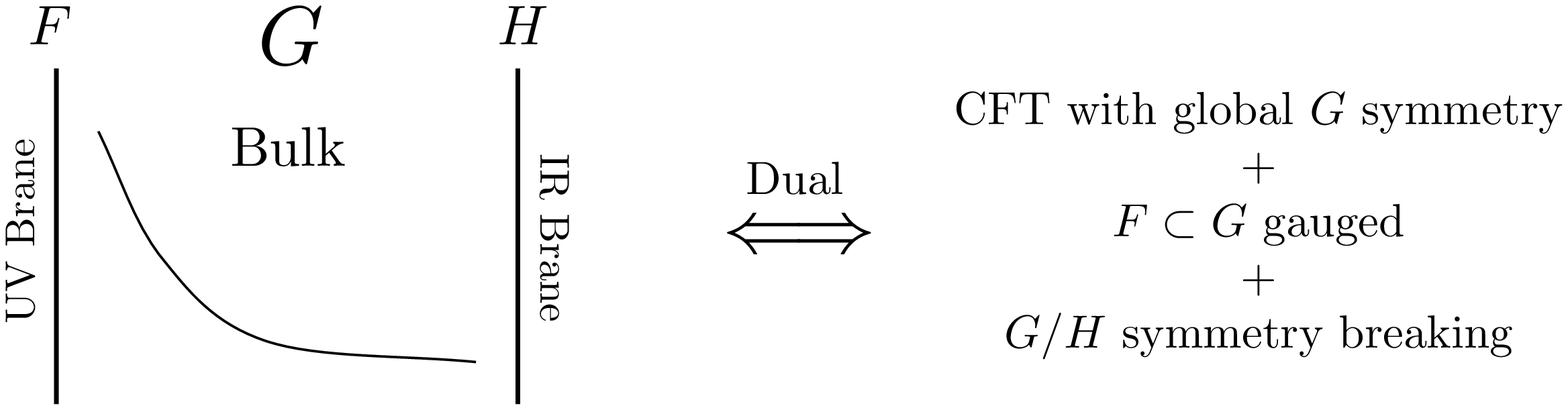}
\end{tabular}
\ee
This construction was studied in the context of the littlest Higgs in \cite{Thaler:2005en}.  In this paper, we take the obvious next step and deconstruct the warped dimension \cite{Arkani-Hamed:2001ca}.  The link fields in the moose are the Wilson lines constructed out of $A_5$, and the warp factor is reflected in the different decay constants on the links \cite{Randall:2002qr}:
\beq{manysitemoose}
\begin{tabular}{c}
\xymatrix@R=.4pc@C=1.4pc{\mathrm{Global:} & G && G &  & G && G \\
& *=<20pt>[o][F]{} \doublerightxyarrow && *=<20pt>[o][F]{} \rightxyarrow & *=<20pt>[o]{\cdots} \rightxyarrow & *=<20pt>[o][F]{} \doublerightxyarrow && *=<20pt>[o][F]{} \\ \mathrm{Gauged:} & F && G &  & G & & H}
\end{tabular}
\ee
Going to the extreme where we only introduce sites corresponding the UV and IR branes, we arrive at a moose diagram which at low energies is supposed to describe a $G/H$ nonlinear sigma model with $F \subset G$ gauged:
\beq{uvirmoose}
\begin{tabular}{c}
\xymatrix@R=.4pc@C=1.4pc{\mathrm{Global:} & G && G \\
& *=<20pt>[o][F]{} \doublerightxyarrow^{\mbox{\raisebox{1.5ex}{$\xi$}}} && *=<20pt>[o][F]{} \\ \mathrm{Gauged:} &F &&H&}
\end{tabular}
\ee
In this AdS/CFT-inspired $G/H$ moose, the subgroup $H$ (which was a global symmetry from the low energy perspective) has become a gauge symmetry.  

Now, this construction of a $G/H$ nonlinear sigma model is quite well-known and is referred to in the literature as Hidden Local Symmetry (HLS) (see \cite{Bando:1987br} for a review).  The link field $\xi$ is identified with the Goldstone matrix in the CCWZ prescription \cite{Coleman:1969sm,Callan:1969sn}, and the $H$ gauge bosons are interpreted as auxiliary fields.   In appendix \ref{sec:warp}, we show how to generate the $G/H$ nonlinear sigma model by explicitly integrating out the bulk of AdS$_5$, but starting from the $G/H$ moose in \eq{uvirmoose}, it is trivial to reproduce the CCWZ results in the spirit of HLS.  

The effective lagrangian for $\xi = e^{i \Pi/ f}$ is
\beq{mooseeft}
\mathcal{L} = -\frac{1}{2g_F^2}\tr F_{\mu\nu}^2 - \frac{1}{2g_H^2}\tr H_{\mu\nu}^2 + f^2 \tr |D_\mu \xi|^2, \qquad D_\mu \xi = \partial_\mu \xi +i F_\mu \xi -i \xi H_\mu,
\ee
where $\Pi \in G$ is the Goldstone matrix, $f$ is the Goldstone decay constant, and $F_\mu$ and $H_\mu$ are the gauge fields associated with the $F$ and $H$ gauge groups.  We can go to a unitary gauge where $\Pi \in G/(F \cup H)$.  At tree level, the spectrum consists of the massless Goldstones from $\xi$, the massless gauge bosons in $F \cap H$, and the massive gauge bosons in $F \cup H$. 

In the limit $g_H \rightarrow \infty$, the massive gauge bosons in $H$ decouple, and we can simply integrate out $H_\mu$.  For convenience, we decompose the object $\xi^\dagger (\partial_\mu + i F_\mu) \xi$ into elements of $H$ and $G/H$:
\be
\label{vpdecomp}
\xi^\dagger (\partial_\mu + i F_\mu) \xi \equiv v_\mu  + p_\mu, \qquad v_\mu \in H, \qquad p_\mu \in G/H.
\ee
To leading order, the $H_\mu$ equation of motion sets
\be
i H_\mu \equiv v_\mu.
\ee
Plugging this value of $H_\mu$ into \eq{mooseeft} in the $g_H \rightarrow \infty$ limit:
\beq{ccwzlagrange}
\mathcal{L}=  -\frac{1}{2g_F^2}\tr F_{\mu\nu}^2 + f^2 \tr p^\mu p^\dagger_\mu,
\ee
which is precisely the CCWZ phenomenological lagrangian.  Similarly, if a fermion $\psi$ is charged under $H$, $\psi \rightarrow h \psi$, its kinetic term before and after integrating out $H_\mu$ is
\beq{fermionvmucoupling}
\mathcal{L}_\psi = \bar{\psi} i \bar{\sigma}^\mu (\partial_\mu + i H_\mu) \psi, \qquad \mathcal{L}_\psi = \bar{\psi} i \bar{\sigma}^\mu (\partial_\mu + v_\mu) \psi + \frac{1}{f^2} \left(\mbox{four-fermion operators}\right),
\ee
in agreement with CCWZ and na\"{\i}ve dimensional analysis (NDA) \cite{Manohar:1983md,Georgi:1986kr} for the size of four fermion operators.  More generally, for finite $g_H$, it is easy to show that the size of higher order interactions from integrating out $H_\mu$ are consistent with NDA with the choice $\Lambda = g_H f$.  Whereas the original CCWZ prescription for arbitrary $G/H$ required some insight to understand the importance of $v_\mu$ and $p_\mu$, the lagrangian in \eq{mooseeft} is trivial, and the CCWZ prescription falls out immediately from integrating out $H_\mu$.

For finite $g_H$, this HLS construction can used as a model for $\rho$ mesons in confining theories \cite{Bando:1984ej} (more precisely, $\rho$ mesons in the ``vector limit'' \cite{Georgi:1989xy}; see also \cite{Harada:2003jx}).  Indeed, in the AdS picture, the $H$ gauge symmetry on the IR brane is the ``gauge symmetry'' associated with light spin-1 KK modes which are holographically identified as CFT resonances.  It has been observed that repeated applications of HLS as a model of higher $\rho$-like resonances can reconstruct an extra dimension \cite{Bando:1985rf, Bando:1987ym,Son:2003et,Piai:2004yb}, and in this light, the CCWZ prescription (as interpreted through the $G/H$ moose) could be seen as the first hint that confining theories holographically generate an extra dimension.  

In the context of little Higgs theories, the $G/H$ moose was implicitly studied in \cite{Piai:2004yb} to try to understand the UV sensitive parameters in the $SU(5)/SO(5)$ littlest Higgs, and they found that by including a $SO(5)$s worth of $\rho$ mesons, all the UV sensitive parameters in the original littlest Higgs became finite to one-loop order.  (See \cite{Harada:2003xa} in the context of ordinary QCD.)  This is easy to understand from the point of view of the $G/H$ moose.  When $g_H = 0$, all of the Goldstones in $\xi$ are exact, so the gauge coupling $g_H$ acts like a spurion for the soft breaking of $SU(5)$.  In the gauge sector of the littlest Higgs, the Higgs potential is now controlled by three spurions instead of just two, so by collective breaking there can be no one-loop quadratically divergent contributions to the Higgs potential.  

Similarly, by extending the moose as in \eq{manysitemoose} to approximate a tower of spin-1 resonances, the Higgs potential can be made arbitrarily well behaved, because the gauge coupling on each site now acts like a spurion for $SU(5)$ breaking \cite{Arkani-Hamed:2001nc}.  This feature has been used to calculate the Higgs potential in AdS$_5$ models where the Higgs is a holographic pseudo-Goldstone boson \cite{Contino:2003ve,Agashe:2004rs,Thaler:2005en}, and as we will discuss in the section \ref{sec:four}, there is always a notion of collective breaking in any holographic Higgs model because both $g_F$ and $g_H$ must be turned on for the Goldstones in $\xi$ to acquire a radiative potential.

So to date, HLS has been a way to discuss the spin-1 phenomenology of confining theories.  In this paper, we observe that the $H$ gauge symmetry can also survive deep into the ultraviolet!  In the case that $G = SU(N)$, we can write down a simple UV completion for the moose diagram in \eq{uvirmoose}.  (The links now correspond to fermions.)
\be
\begin{tabular}{c}
\xymatrix@R=.4pc@C=1.4pc{\mathrm{Global:} & SU(N)_L && SU(N)_R \\
& *=<20pt>[o][F]{} \rightxyarrow^{\mbox{\raisebox{1.5ex}{$\psi$}}} & *=<12pt>[o][F=]{} \rightxyarrow^{\mbox{\raisebox{1.5ex}{$\psi^c$}}} & *=<20pt>[o][F]{} \\ \mathrm{Gauged:} &F &SU(N_c)&H}
\end{tabular}
\ee
We recognize this as the moose diagram for QCD with $N$ flavors, where some of the flavor symmetries have been gauged.  When $SU(N_c)$ confines, the fermion condensate $\langle \psi \psi^c \rangle$ will become the link field $\xi$.  Specializing to the littlest Higgs (with only one hypercharge generator):
\beq{littlehiggsmoose}
\begin{tabular}{c}
\xymatrix@R=.4pc@C=1.4pc{\mathrm{Global:} & SU(5)_L && SU(5)_R \\
& *=<20pt>[o][F]{} \rightxyarrow^{\mbox{\raisebox{1.5ex}{$\psi$}}} & *=<12pt>[o][F=]{} \rightxyarrow^{\mbox{\raisebox{1.5ex}{$\psi^c$}}} & *=<20pt>[o][F]{} \\ \mathrm{Gauged:} &SU(2)^2 \times U(1)_Y &SU(N_c)&SO(5)}
\end{tabular}
\ee
Apparently, the $SU(5)/SO(5)$ littlest Higgs is contained in ordinary QCD with five flavors with the appropriate gauging of the flavor symmetries.  In other words, this theory \emph{is} technicolor, albeit with different gauge groups.  Below the confinement scale, the $SO(5)$ gauge bosons will become heavy via chiral symmetry breaking, and assuming the $SO(5)$ gauge coupling is large enough, we can integrate them out to generate a $SU(5)/SO(5)$ nonlinear sigma model.

In other words, this ``littlest technicolor'' model has exactly the same low energy phenomenology as the original littlest Higgs, despite the fact that we have introduced a new gauge symmetry.  Of course, now that we understand that little Higgs theories could come from technicolor-like theories, we inevitability encounter some of the same tensions from technicolor, such as how to generate fermion Yukawa couplings without introducing new sources of flavor violation \cite{Eichten:1979ah,Dimopoulos:1979es}.  (We discuss some aspects of fermions in appendix \ref{sec:fermions}.)  Also, as presented, the moose in \eq{littlehiggsmoose} has anomalies  associated with the $F$ gauge group (but no anomalies that involve $H$ because $\psi^c$ transforms as a real representation of $SO(5)$), so one would have to hope that some of standard model fermions act as spectators to cancel these anomalies.  On the other hand, it is remarkable that such a simple UV completion exists.

Before going on to discuss little technicolor in the context of the simple group little Higgs, we remark that our attitude to spin-1 fields is similar in spirit to the Abbott-Farhi model of electroweak gauge bosons \cite{Abbott:1981re,Abbott:1981yg}.  Ignoring hypercharge, the moose diagram for a massive $W$ boson with a custodial $SU(2)$ symmetry is: 
\beq{abbotfarhistart}
\begin{tabular}{c}
\xymatrix@R=.4pc@C=1.4pc{\mathrm{Global:} & SU(2) && SU(2) \\
& *=<20pt>[o][F]{} \doublerightxyarrow && *=<20pt>[o][F]{} \\ \mathrm{Gauged:} &SU(2) && &}
\end{tabular}
\ee
If the only thing we knew about the $W$ boson was that it was a massive spin-1 field, then we could imagine many different UV completions for \eq{abbotfarhistart} that would unitarize $W$ longitudinal mode scattering.   In the standard model, the link field is assumed to be UV completed into a linear sigma field.  In technicolor, the link field comes from a fermion condensate.  We could also imagine this moose diagram as a picture of a slice of AdS$_5$.  If we let the gauged $SU(2)$ site be the UV brane,
\be
\begin{tabular}{c|c}
& Gauge Symmetry\\
\hline
UV Brane &  $SU(2)$ \\
Bulk & $SU(2)$ \\
IR Brane &  $\emptyset$
\end{tabular}
\qquad
\mathop{\Longleftrightarrow}^{\raisebox{.4ex}{Dual}}
\qquad
\begin{tabular}{c}
CFT with Gauged $SU(2)$ \\
$+$\\
$SU(2)/ \emptyset$ Symmetry Breaking
\end{tabular}
\ee
then the $W$ gauge boson gets a mass via spontaneous symmetry breaking, \`{a} la technicolor.   Alternatively, we could reverse the roles of the UV and IR branes,
\be
\begin{tabular}{c|c}
& Gauge Symmetry\\
\hline
UV Brane &  $\emptyset$ \\
Bulk & $SU(2)$ \\
IR Brane &  $SU(2)$
\end{tabular}
\qquad
\mathop{\Longleftrightarrow}^{\raisebox{.4ex}{Dual}}
\qquad
\begin{tabular}{c}
CFT with Global $SU(2)$ \\
$+$\\
No Symmetry Breaking
\end{tabular}
\ee
in which case the $W$ boson is a $\rho$-like meson of a confining theory, which is the Abbott-Farhi model.   Of course, the detailed properties of an ultraviolet gauge boson and an composite spin-1 meson are very different, and we can distinguish between Abbott-Farhi, technicolor, and the standard model through precision electroweak tests.  But given a generic massive spin-1 field, we have no \emph{a priori} reason to label it a $W$-like vs. a $\rho$-like state.  As we will see for the simple group little Higgs, by allowing ourselves maximal flexibility in relabeling the spin-1 spectrum, we can come up with many UV complete models with the same low energy physics.  

\section{Four Views of the Simple Group Little Higgs}
\label{sec:four}

We have seen that the $H_\mu$ field can be an auxiliary field for deriving CCWZ or a model of the $\rho$-like mesons in HLS.  The key point of this paper is that it is entirely consistent for the $H_\mu$ field to be a real gauge field that survives deep into the ultraviolet.  The simple group little Higgs model \cite{Kaplan:2003uc} is an ideal laboratory to study this possibility, because the symmetry structure of the theory makes it possible to imagine many different UV completions.  In essence, we will show that the distinction between $W'$-like gauge bosons and $\rho$-like mesons can be blurred in these models.  Moreover, we will understand that  holographic composite Higgs models are actually little Higgs theories in the sense that there is always a meaning to collective breaking.

The simple group little Higgs is based on an $(SU(3)/SU(2))^2$ symmetry breaking pattern where the diagonal $SU(3)$ symmetry is gauged.  (We are ignoring hypercharge for simplicity.)  Using the AdS/CFT correspondence as in \eq{duality}, we can easily construct an AdS$_5$ model to recover the low energy physics of the simple group theory.  The brane and bulk gauge symmetries are:
\beq{adssimplegroup}
\begin{tabular}{c|c}
& Gauge Symmetry\\
\hline
UV Brane &  $SU(3)_V$ \\
Bulk & $SU(3)_1 \times SU(3)_2$ \\
IR Brane &  $SU(2)_1 \times SU(2)_2$
\end{tabular}
\ee
At low energies, there are the massless $SU(2)_{EW}$ gauge bosons and an $SU(3)/SU(2)$'s worth of Goldstones, which contain an electroweak doublet (the Higgs) and an electroweak singlet.  

Like all little Higgs theories, the simple group theory exhibits collective breaking in that no single interaction breaks the global symmetries that protect the Higgs potential, so the Higgs mass is not quadratically sensitive to the cutoff (\emph{i.e.}\ the confinement scale).   It is easy to see how collective breaking works in AdS space in terms of boundary conditions on the UV and IR branes \cite{Thaler:2005en}.  When the gauge boson boundary conditions are Neumann on either brane, then we can go to $A_5 = 0$ gauge, and all the Goldstones are exact (\emph{i.e.}\ they are all eaten).  Similarly, when the gauge boson boundary conditions are Dirichlet on either brane, then there are no massless gauge fields for the Goldstones to mix with, so all the Goldstones are exact.  Only when different components of the bulk gauge field have different boundary conditions on each brane is it possible to have pseudo-Goldstone bosons.   In the simple group theory, there are three interactions that have to be turned on for the symmetries that protect the Higgs to be broken:  $SU(3)_V$ has to be gauged on the UV brane, $SU(3)_1$ has to be reduced to $SU(2)_1$ on the IR brane, and similarly for $SU(3)_2$.

The lightest KK modes with Neumann boundary conditions on the IR brane are holographically identified with the $\rho$-like mesons of the confining CFT.  We will now perform the two-site deconstruction of the AdS$_5$ model in \eq{adssimplegroup} to show that these $\rho$ mesons can be interpreted as $W'$ gauge bosons in a theory of two copies of QCD with $N_f = 3$.  Following the example of \eq{uvirmoose}, the relevant moose diagram is:
\be
\begin{tabular}{c}
\xymatrix@R=.4pc@C=1.4pc{\mathrm{Global:} & SU(3)^2 && SU(3)^2& \\
& *=<20pt>[o][F]{} \doublerightxyarrow && *=<20pt>[o][F]{} \\ \mathrm{Gauged:} &SU(3)_V &&SU(2)^2&}
\end{tabular}
\ee
Because $SU(3)^2$ is a direct product space, we can decompose this moose into two pieces:
\beq{vectorlimitmoose}
\begin{tabular}{c}
\xy
\xymatrix@R=.4pc@C=1.4pc{\mathrm{Global:} & SU(3) && SU(3) & SU(3) && SU(3)& \\
& *=<20pt>[o][F]{} \doublerightxyarrow && *=<20pt>[o][F]{} & *=<20pt>[o][F]{}      \doublerightxyarrow && *=<20pt>[o][F]{} \\ \mathrm{Gauged:} &SU(2) &&\ar @{} [r] |{\mbox{\raisebox{0.0ex}{$SU(3)_V$}}} &&&  SU(2)&  \save "2,4"-(5,5);"2,5"+(5,5) **\frm{--}  \restore} 
\endxy
\end{tabular}
\ee
The $SU(3)_V$ gauge symmetry breaks the approximate $SU(3)$ global symmetries on the boxed sites, so we might as well combine those two-sites into a single site.  We can now UV complete the link fields into fermion condensates from two $SU(N_c)$ confining theories (arrows now correspond to fermions):
\be
\begin{tabular}{c}
\xymatrix@R=.4pc@C=1.4pc{\mathrm{Global:} & SU(3) && SU(3) && SU(3) & \\
& *=<20pt>[o][F]{} \rightxyarrow & *=<12pt>[o][F=]{} \rightxyarrow & *=<20pt>[o][F]{} \rightxyarrow& *=<12pt>[o][F=]{} \rightxyarrow&  *=<20pt>[o][F]{}  \\ \mathrm{Gauged:} &SU(2) &SU(N_c)&SU(3) & SU(N_c) & SU(2) &}
\end{tabular}
\ee
We see that the low energy degrees of freedom in the simple group theory can arise from two copies of QCD with $N_f = 3$, where some of the flavor symmetries have been gauged.  Indeed, the $SU(2)^2$ gauge symmetries which had been associated with the holographic $\rho$-like mesons in the AdS$_5$ model are now ultraviolet gauge symmetries associated with $W$ and $W'$ gauge fields.  If we had deconstructed the original AdS$_5$ model with three sites instead of two, we could do the exact same construction to see that the simple group theory could be embedded into four copies of QCD, each with $N_f = 3$ and the appropriate flavor symmetries gauged.  It is obvious from this construction that any $(SU(N)/H)^n$ little Higgs theory where some subgroup of $SU(N)^n$ is gauged can arise from $n$ copies of QCD with $N$ flavors with the appropriate gauging of the flavor symmetries.  

There is an even more interesting UV completion of the moose in \eq{vectorlimitmoose} inspired by the HLS model of the $\rho$ meson in ordinary QCD \cite{Bando:1984ej}.  As postulated by \cite{Georgi:1989xy}, there may be a ``vector limit'' of QCD (possibility the large $N_c$ limit) where the $\rho$ meson is anomalously light compared to other QCD resonances.  Above the confinement scale, the QCD moose with $N_f$ fermion flavors is:
\be
\begin{tabular}{c}
\xymatrix@R=.4pc@C=1.4pc{\mathrm{Global:} & SU(N_f)_L && SU(N_f)_R & \\
& *=<20pt>[o][F]{} \rightxyarrow & *=<12pt>[o][F=]{} \rightxyarrow & *=<20pt>[o][F]{}  \\ \mathrm{Gauged:} & &SU(N_c)&}
\end{tabular}
\ee
When we describe only the pions of QCD, the low energy description is in terms of the nonlinear sigma model moose: 
\be
\begin{tabular}{c}
\xymatrix@R=.4pc@C=1.4pc{\mathrm{Global:} & SU(N_f)_L && SU(N_f)_R & \\
& *=<20pt>[o][F]{} \doublerightxyarrow & & *=<20pt>[o][F]{}}
\end{tabular}
\vspace{.15in}
\ee
The hypothesis of  \cite{Georgi:1989xy} is that the effective description of the pions and $\rho$ mesons in the vector limit of QCD should be:
\beq{vectorlimittwo}
\begin{tabular}{c}
\xy
\xymatrix@R=.4pc@C=1.4pc{\mathrm{Global:} & SU(N_f)_L && SU(N_f)_L' & SU(N_f)_R' && SU(N_f)_R \\
& *=<20pt>[o][F]{} \doublerightxyarrow && *=<20pt>[o][F]{} & *=<20pt>[o][F]{}      \doublerightxyarrow && *=<20pt>[o][F]{} \\ \mathrm{Gauged:} & &&\ar @{} [r] |{\mbox{\raisebox{0.0ex}{$SU(N_f)_V$}}} &&  \save "2,4"-(5,5);"2,5"+(5,5) **\frm{--}  \restore} 
\endxy
\end{tabular}
\ee
where $SU(N_f)_V$ is the gauge symmetry associated with the $\rho$ mesons.  The point of the vector limit is that as the $SU(N_f)_V$ gauge coupling is taken to zero, there could be an enhanced $SU(N_f)^4$ ``flavor'' symmetry in QCD which would allow the longitudinal components of $\rho$ to be the exact chiral partners of the pions.

In fact, the AdS$_5$ metaphor for large $N_c$ QCD with $N_f$ flavors (and no flavor symmetries gauged) realizes this scenario explicitly (see \emph{e.g.}~\cite{Erlich:2005qh} when $N_c = 3$).  The gauge symmetries in the AdS$_5$ metaphor are:
\be
\begin{tabular}{c|c}
& Gauge Symmetry\\
\hline
UV Brane &  $\emptyset$ \\
Bulk & $SU(N_f)_L \times SU(N_f)_R$ \\
IR Brane &  $SU(N_f)_V$
\end{tabular}
\ee
and when we deconstruct the AdS space, we indeed generate the moose in \eq{vectorlimittwo}, and the higher AdS KK states are well separated from the $\rho$ meson.  What is amusing is that in ordinary QCD, $SU(N_f)_V$ is the gauge symmetry associated with the $\rho$ meson, whereas in the original inspiration for the simple group little Higgs, $SU(3)_V$ is a gauge symmetry associated with an ultraviolet gauge boson.  From the point of view of low energy degrees of freedom, however, the mooses in \eqs{vectorlimitmoose}{vectorlimittwo} are identical if we set $N_f = 3$ and gauge an $SU(2)^2$ subgroup of $SU(3)_L \times SU(3)_R$.  So we see that the simple group theory can arise from one copy of QCD with 3 flavors in the vector limit!  What was an ultraviolet $W$ gauge boson in the original theory is a $\rho$ meson in this QCD model, but by construction, both theories have the same low energy physics.

From ordinary QCD where $N_f = N_c = 3$, we know that $g_\rho = m_\rho / f_\pi \sim 8$, which is too large to be the perturbative $SU(3)_V$ gauge coupling in the simple group Higgs.  The question is whether we can decrease $g_\rho$ by going to a large $N_c$ theory.  Conventionally, $m_\rho$ is thought to be fixed as $N_c$ increases, so by using the $\sqrt{N_c}$ scaling of $f_\pi$ in large $N$ theories \cite{'tHooft:1973jz}, we would predict that $g_\rho$ scales as $1/\sqrt{N_c}$.  However, if $m_\rho$ is fixed as $N_c$ increases, then there is no reason to expect $m_\rho$ to be well separated from ``QCD string'' states unless we are in a conformal window as in AdS/CFT or if the value of $N_f$ puts us close to the chiral symmetry breaking phase transition \cite{Harada:2000kb}.  In other words, while $g_\rho$ would be small at large $N_c$, the effective theory would not be well described by the moose in \eq{vectorlimittwo} because of pollution from higher QCD resonances.  A hypothesis presented in \cite{Georgi:1989xy} is that the vector limit could arise if $g_\rho$ decreases \emph{faster} than $1/\sqrt{N_c}$, such that $m_\rho$ goes to zero in the large $N_c$ limit.  This would guarantee a parametric separation between the $\rho$ mesons and higher QCD resonances, justifying the moose description in \eq{vectorlimittwo}.

Finally, we can interpret \eq{vectorlimitmoose} as a picture of an AdS$_5$ space where the left-most site is associated with the UV brane and the right-most site is associated with the IR brane.
\be
\begin{tabular}{c|c}
& Gauge Symmetry\\
\hline
UV Brane &  $SU(2)$ \\
Bulk & $SU(3)$ \\
IR Brane &  $SU(2)$
\end{tabular}
\ee
Modulo hypercharge, this symmetry pattern is exactly the symmetry pattern of the original realization of the Higgs as a holographic pseudo-Goldstone boson \cite{Contino:2003ve}.  Like the $(SU(3)/SU(2))^2$ simple group, the minimal model in \cite{Contino:2003ve} does not have a natural mechanism for generating a large Higgs quartic coupling, and this is to be expected because both theories have the same low energy degrees of freedom.  So without additional dynamics, neither model can successfully trigger electroweak symmetry breaking.  The $(SU(4)/SU(3))^4$ simple group \cite{Kaplan:2003uc} does have a way to generate a large Higgs quartic coupling, so an AdS$_5$ model with bulk $SU(4)$ or $SU(4)^2$ gauge bosons might be a good starting point to construct a viable model of a composite Higgs boson, assuming there is an AdS$_5$ analog of the interactions that generate the low energy quartic coupling.  (See, however, section \ref{sec:fvsh}.)  

But whereas we call the simple group theory a ``little Higgs'' theory, generic implementations of the Higgs as a holographic pseudo-Goldstone boson are not referred to as little Higgs theories in the literature.   As we mentioned at the beginning of this section, though, \emph{any} AdS$_5$ model  with an electroweak doublet Goldstone is a little Higgs theory because there is always a notion of collective breaking, namely collectively choosing boundary conditions on the UV and IR branes.  In terms of the two-site moose from \eq{uvirmoose}, collective breaking means that both the $g_F$ and $g_H$ gauge couplings must be turned on in order for the uneaten Goldstones to acquire a radiative potential, and when we interpret the $H_\mu$ gauge fields as $\rho$-like mesons from a confining theory, there is a useful notion of collective breaking to the extent that the $\rho$s are light (\emph{i.e.}\ the $H_\mu$ gauge coupling is small, or equivalently, we are in the vector limit of the confining theory).

In other words, what makes an interesting theory of a composite Higgs boson is not collective breaking \emph{per se}, but whether the Higgs potential can successfully trigger electroweak symmetry breaking while maintaining a large enough hierarchy between the confinement scale and the electroweak scale to avoid precision electroweak constraints.  Any old symmetry breaking pattern that yields an electroweak doublet Goldstone will be a little Higgs theory if there are light $\rho$ mesons in the spectrum, but only very special theories can generate a light Higgs with a large quartic coupling, the necessary ingredients for a successful composite Higgs model.  In the case that the approximate global symmetries are $SU(N)$s, we can use the technique described in this section to UV complete $(SU(N)/H)^n$ little Higgs models into technicolor.  The littlest Higgs satisfies both these criteria, and in appendix \ref{sec:fermions} we sketch how to implement the top sector for the moose in \eq{littlehiggsmoose}.

\section{Holographic Pseudo-Goldstones as Little Higgses}
\label{sec:holo}

We have argued that holographic composite Higgs models are little Higgs theories to the extent that we can define a notion of collective breaking in AdS space, and this fact is obvious in light of the AdS/CFT correspondence.  If the gauge symmetry $F$ on the UV brane is equal to $G$, then all of the flavor symmetries in the dual CFT are gauged so there are no uneaten Goldstone modes.  If the gauge symmetry $H$ on the IR brane is equal to $G$, then there is no spontaneous symmetry breaking in the dual theory and therefore no Goldstones.  (Analogous arguments hold when $F$ or $H$ are empty.)  In other words, collective breaking in AdS space reduces to the condition for the existence of pseudo-Goldstone bosons in the dual CFT \cite{Georgi:1975tz}.

Inspired by the $G/H$ moose, there is another way that we can see collective breaking in models with holographic pseudo-Goldstone bosons.  By ``integrating in'' the IR brane, we will see that these models have identical low energy physics to brane-localized little technicolor theories, and the radiative Higgs potential can be made arbitrarily finite without ever having to discuss the details of the warped dimension.  In effect, the finiteness of the Higgs potential has to do only with a small slice of AdS space, and if we are interested only in studying a calculable composite Higgs model, then we need not invoke the full machinery of AdS/CFT.  We will be able to turn holographic $\rho$ mesons into brane-localized $W'$ gauge bosons, and the low energy physics will be shielded from bulk dynamics.

While the original inspiration for the $G/H$ moose came from AdS/CFT, we show in appendix \ref{sec:warp} that a $G/H$ nonlinear sigma model can arise from any extra dimension with boundaries, including flat space.  One advantage of AdS space over flat space is that integrating out heavy modes is almost exactly the same as integrating out sites of the deconstructed moose.  If the lattice spacing is much larger than the AdS length, then from \eq{unitarygaugeAdSaction}, we see that integrating out the UV brane site is almost exactly the same as integrating out the heaviest mode in the spectrum.  In continuum language, we can move the UV brane while keeping the low energy physics fixed by allowing couplings on the UV brane to run, and this running is holographically dual to renormalization group running in the CFT.  In flat space, integrating out heavy modes involves all of the sites, and therefore introduces interactions between the remaining sites that are non-local in theory space.  In AdS space, the induced non-local interactions are exponentially suppressed and can be ignored to an excellent approximation.  

For the same reason, if we want to move the IR brane while keeping the low energy physics fixed, we have to introduce new degrees of freedom on the IR brane.   This takes into account the fact that by moving the IR brane, we are integrating out a light degree of freedom which has $\mathcal{O}(1)$ overlap with the IR brane site, so we have to integrate this degree of freedom back into the spectrum.  By deconstructing AdS as in appendix \ref{sec:warp}, it is easy to see that the following theories have identical low energy physics:
\medskip
\beq{littlesimilar}
\begin{tabular}{c}
\includegraphics[scale=0.50]{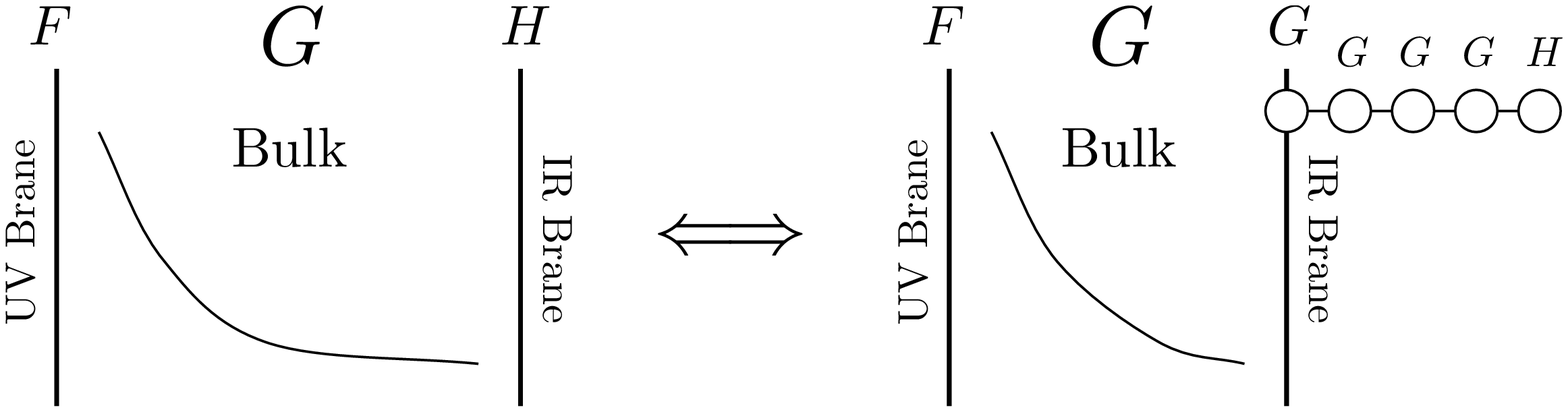}
\end{tabular}
\ee
where we have placed the $N$ site moose
\beq{nsitemoose}
 \begin{tabular}{c}
\xymatrix@R=.4pc@C=1.4pc{G_{\rm global} & G_{\rm gauged} && G_{\rm gauged}  & H_{\rm gauged} \\
*=<20pt>[o][F]{} \rightxyarrow  &  *=<20pt>[o][F]{} \rightxyarrow &\cdots \rightxyarrow &  *=<20pt>[o][F]{}\rightxyarrow & *=<20pt>[o][F]{}
 }
\end{tabular}
\vspace{.15in}
\ee
on the IR brane and identified $G_{\rm global}$ with the bulk gauge symmetry.   In order to reproduce the desired warp factor as in \eq{deconAdSaction}, we have to give an exponential profile to the decay constants on the $N$ site moose.  If the number of sites is small, then we may not even need to UV complete the link fields on the brane because unitarity would not be saturated until well beyond the local quantum gravity scale.  Alternatively, the link fields could be embedded into linear sigma models without any considerable fine-tuning, or we could UV complete \eq{nsitemoose} into brane-localized little technicolor.  

What is interesting about integrating in the IR brane is that it makes it clear that only a small slice of AdS space is necessary to get a finite radiative potential for the holographic pseudo-Goldstones.  In the case of the $SU(5)/SO(5)$ littlest Higgs, the one-loop radiative Higgs potential for an AdS space with
\be
F = SU(2)^2 \times U(1), \qquad G = SU(5), \qquad H = SO(5),
\ee
was calculated in \cite{Thaler:2005en}.  However, by \eq{littlesimilar}, we could just as well transition to the analysis of \cite{Piai:2004yb} for the moose in \eq{nsitemoose} with an $F$ subgroup of the $G_{\rm global}$ site gauged.  In other words, for purposes of understanding the radiative potential for a composite Higgs, a holographic composite model and an $N$ site moose localized on the IR brane will give nearly identical results, and the results can be made arbitrarily similar by adjusting the decay constants and gauge couplings on the $N$ site moose to match the holographic resonances of the initial CFT.

For concreteness, in the original littlest Higgs \cite{Arkani-Hamed:2002qy}, the radiative Higgs quartic coupling from the gauge sector was quadratically sensitive to the cutoff:
\be
\lambda \sim \frac{\Lambda^2}{(4\pi f)^2} g_{EW}^2,
\ee
where $g_{EW}$ is the low energy $SU(2)_{EW}$ gauge coupling.  In the holographic model of \cite{Thaler:2005en}, the Higgs potential was finite and parametrically scaled as
\be
\lambda \sim \frac{1}{N_c}g_{EW}^2.
\ee
In the littlest Higgs with a composite $\rho$ meson modeled by HLS (\emph{i.e.}\ \eq{nsitemoose} with two sites) \cite{Piai:2004yb}, the Higgs potential was logarithmically sensitive to the scale of the second $\rho$-like resonances, but setting these logarithms to one, the Higgs quartic scaled as:
\be
\lambda \sim \frac{m_\rho^2}{(4\pi f)^2} g_{EW}^2.
\ee
(We could remove all logarithmic sensitivity by transitioning to a three site moose.)  If we make the identification  $\Lambda \sim m_\rho \sim 4 \pi f / \sqrt{N_c}$ \cite{'tHooft:1973jz}, then these results scale in the same way as expected.  

More importantly, once we believe that \eq{littlesimilar} accurately describes the same low energy physics, the analysis of \cite{Thaler:2005en} and \cite{Piai:2004yb} for the one-loop radiative Higgs potential will be nearly identical once we match the spectra of the two theories.  Because we have moved the IR brane in, pollution from bulk gauge boson KK modes as well as 5D quantum gravity effects occur at a much higher scale, so corrections to the radiative potential from bulk dynamics or from the details of how the brane-localized moose is UV completion are very suppressed.

In other words, the statement that holographic pseudo-Goldstones are little Higgses is another way of saying that locality in AdS space can me mimicked by locality in theory space.  Indeed, by integrating in the IR brane, locality in AdS space \emph{becomes} locality in theory space.  In little technicolor models, collective breaking is manifest because the gauge coupling on each site has to be turned on for the Higgs to pick up a radiative potential.  The analysis of \cite{Piai:2004yb} further guarantees us that at the one-loop level, only a few sites are necessary on the IR brane to get finite and calculable radiative corrections without ever needing to understand bulk dynamics. In appendix \ref{sec:bc} we consider the reverse situation, and show how all information about how $G$ is reduced to $H$ on the IR brane can be absorbed into small perturbations of the warp factor near the IR brane.

\section{Vacuum Alignment}
\label{sec:fvsh}

One challenge to constructing viable little Higgs theories is vacuum alignment.  In the $SU(6)/Sp(6)$ little Higgs \cite{Low:2002ws}, there is a gauged $SU(2)^2$ subgroup of $SU(6)$, and it is essential that the breaking of $SU(6) \rightarrow Sp(6)$ also breaks $SU(2)^2 \rightarrow SU(2)$.  Similarly, in the $(SU(4)/SU(3))^4$ theory \cite{Kaplan:2003uc}, pairs of the $SU(3)$ subgroups must be misaligned when the diagonal $SU(4)$ gauge symmetry is turned on in order to have a light Higgs doublet with a large quartic coupling.  Therefore, an ultraviolet completion of a little Higgs theory must not only break $G \rightarrow H$, but when an $F$ subgroup of $G$ is gauged, the relative orientation of $F$ and $H$ (and the various subgroups of $H$) must be radiatively stable. 

In little technicolor, we are specifying the UV completion of \eq{intromoose} where QCD dynamics guarantees that $SU(N)_L \times SU(N)_R$ will break to the diagonal $SU(N)$ as long as we are on the correct side of the chiral symmetry breaking phase transition.  Once we choose a generator basis for $SU(N)_L \times SU(N)_R$, we can of course gauge any subgroup $F \times H$ we wish.   However, there is no guarantee that in this basis the fermion condensate $\langle \psi \psi^c \rangle$ will point in the direction of the identity.   If we assume that the $SU(N)_L \times SU(N)_R$ flavor symmetry is exact above $\Lambda_{QCD}$, then the orientation of $\langle \psi \psi^c \rangle$ (and hence the relative orientation of $F$ and $H$) is determined solely by radiative corrections.  

Vacuum alignment for the little technicolor version of $SU(6)/Sp(6)$ theory was implicitly studied in \cite{Piai:2004yb} where it was found that gauge boson loops do not favor $SU(2)^2 \rightarrow SU(2)$ breaking.  The little technicolor version of the $(SU(4)/SU(3))^4$ simple group theory has the following low energy moose:
\beq{fourtothreemoose}
\begin{tabular}{l | l l}
& Global & Gauged \\
\hline
$A$ &$SU(4)$& $SU(4)$\\
$B$ & $SU(4)$& $SU(3)$\\
$C$ & $SU(4)$& $SU(3)'$
\end{tabular}
\qquad \qquad
\begin{tabular}{c}
\xymatrix@R=.4pc@C=1.4pc{*=<20pt>[o][F]{B} \doublerightdownxyarrow && && *=<20pt>[o][F]{C}\\
&& *=<20pt>[o][F]{A} \doublerightupxyarrow \doublerightdownxyarrow \\
*=<20pt>[o][F]{B} \doublerightupxyarrow   &&&&*=<20pt>[o][F]{C}}
\end{tabular}
\ee
where $SU(3)$ and $SU(3)'$ are misaligned in some suitable basis.  However, it can be shown that the gauge loop Coleman-Weinberg potential \cite{Coleman:1973jx} for the link vevs wants to align $SU(3)$ and $SU(3)'$, agreeing with the general lore that the vacuum points in the direction that yields the largest number of light gauge bosons.  

Of course, it is possible to force a particular vacuum alignment by including operators in the ultraviolet theory that explicitly break the global symmetries.  For example, one could include four fermion operators in the ultraviolet completion of \eq{fourtothreemoose} to force the misalignment of $SU(3)$ and $SU(3)'$ at tree level.   Also, radiative contributions from the fermion sector may affect the choice of vacuum.   The point to keep in mind when constructing ultraviolet theories based on little technicolor is that while there is complete freedom (up to anomaly cancellation) in the choice of $F$ and $H$, one must check that the stable vacuum gives the desired spectrum of light gauge bosons.  In the case of the littlest Higgs in \eq{littlehiggsmoose}, the vacuum indeed aligns in the correct direction \cite{Piai:2004yb}. 

\section{Future Directions}
\label{sec:future}

In this paper, we have explored a beautiful connection between an array of ideas that were already tantalizingly related, but which become even more intertwined in the context of AdS/CFT.   HLS is a model for the $\rho$ mesons in QCD which is known to generate $G/H$ flavor symmetry breaking that is phenomenologically described by CCWZ which is the key to constructing little Higgs theories.  AdS/CFT allows us to explicitly realize the vector limit (\emph{i.e.}\ large $N_c$ or light $\rho$ meson limit) of a confining theory, and once we deconstruct the corresponding AdS$_5$ model, we are free to interpret the link fields either as Wilson lines constructed out of $A_5$, or as fermion condensates from a completely different QCD theory.  In other words, we can interpret the light spin-1 degrees of freedom either as holographic  $\rho$ mesons or as ultraviolet $W'$ gauge fields that acquire a mass via spontaneous symmetry breaking.  Similarly, by integrating in the IR brane in AdS$_5$ models, we saw that holographic $\rho$ mesons can be reinterpreted as brane-localized $W'$ gauge bosons.

Note that there is no deep duality between the holographic CFT and the new QCD theory.  Just like Abbott-Farhi, all we have shown is that two different theories can have the same low energy degrees of freedom, and this is particularly interesting in the context of the little Higgs, where the relevant phenomenology depends almost entirely on the low energy spectrum.  What is a bit surprising (but obvious in retrospect) is that one possible UV completion of an $(SU(N)/H)^n$ little Higgs is $n$ copies of technicolor!  We have known since \cite{Weinberg:1975gm,Susskind:1978ms} that ordinary QCD could give masses to $W$ and $Z$ bosons through spontaneous symmetry breaking; now we know that ordinary QCD can give rise to an electroweak doublet pion which can subsequently break electroweak symmetry.  This allows us (or dooms us, depending on your perspective) to use the full machinery of extended technicolor \cite{Eichten:1979ah,Dimopoulos:1979es} and walking technicolor \cite{Holdom:1981rm} to address phenomenological questions in little Higgs theories (see appendix \ref{sec:fermions}).

What is not obvious is whether general $(G/H)^n$ symmetry breaking patterns can emerge out of moose-like confining theories. A two site moose with global $SU(N)$ sites looks like the pion nonlinear sigma model of QCD, but in order to construct moose diagrams for $SO(N)/H$ theories (such as the $SO(9)/(SO(5)\times SO(4))$ model of \cite{Chang:2003zn}), we would have to find a UV completion for the moose link field:
\be
\begin{tabular}{c}
\xymatrix@R=.4pc@C=1.4pc{\mathrm{Global:} & SO(N) && SO(N) \\
& *=<20pt>[o][F]{} \doublerightxyarrow^{\mbox{\raisebox{1.5ex}{$\xi$}}} && *=<20pt>[o][F]{} \\ \mathrm{Gauged:} & &&H}
\end{tabular}
\quad
\mathop{\Longrightarrow}^{\mbox{?}}
\quad
\begin{tabular}{c}
\xymatrix@R=.4pc@C=1.4pc{SO(N)_1 && SO(N)_2 \\
 *=<20pt>[o][F]{} \rightxyarrow^{\mbox{\raisebox{1.5ex}{$\psi_1$}}} & *=<12pt>[o][F=]{} \rightxyarrow^{\mbox{\raisebox{1.5ex}{$\psi_2$}}} & *=<20pt>[o][F]{} \\  &??? &H}
\end{tabular}
\ee
In order to have $SO(N)$ flavor symmetries, the confining group must have real representations.  To keep the $SO(N)^2$ flavor symmetry from enlarging to an $SO(2N)$ flavor symmetry, the confining group must have two different real representations $\mathbf{R_1}$ and $\mathbf{R_2}$.  Furthermore, for $SO(N)^2$ to spontaneously break to $SO(N)_V$ via confinement, the most attractive condensate must be $\langle \mathbf{R_1} \mathbf{R_2} \rangle$.  Whether these situations can in fact be realized is an open question, though it seems highly unlikely for $G$ to be one of the exceptional groups.  Alternatively, if $G$ is some subgroup of $SU(N)$, we could start with $SU(N)^2/SU(N)$ chiral symmetry breaking and introduce large $SU(N)$-violating (but $G$-preserving) spurions on each of the sites to lift the masses of the unwanted Goldstones. Interestingly, the only constraint on the subgroup $H$ is anomaly cancelation, because we can weakly gauge any global symmetry we wish.

We have seen that the $G/H$ moose in some sense trivializes the CCWZ prescription, and is interesting to wonder whether the same construction might trivialize another consequence of spontaneous symmetry breaking:  the WZW term \cite{Wess:1971yu,Witten:1983tw}.   The WZW term accounts for the fact that anomalies above the confinement scale must match anomalies in the low energy effective theory \cite{'tHooft:1980xb}.  It is well-known that a Chern-Simons (CS) term in the bulk of AdS$_5$ can transmit anomalies from the UV brane to the IR brane \cite{Callan:1984sa}, giving a nice 5D representation of anomaly matching in the holographic CFT.  It is also well-known that the WZW term can be derived by considering a CS term in the interior of a 5-ball, with ordinary 4D space-time on the boundary of the ball \cite{Witten:1983tw,Zumino:1983rz}, though to concisely express the WZW only in terms of phenomenological variables requires the full machinery of differential geometry \cite{Manes:1984gk}.  It is even known how to derive the WZW term in the language of HLS \cite{Wu:1984pv,Fujiwara:1984mp}, so it should be possible to understand the WZW term in term of the $G/H$ moose.  

The difficulty is how to deconstruct the CS term in AdS$_5$, because by design, gauge transformations of the CS term introduce new terms on the UV and IR branes.  Deconstructing a dimension usually involves going to $A_5= 0$ gauge and then reintroducing Wilson line link fields, and in principle, one should track changes in the CS term through this whole procedure.  In the context of QCD, the deconstructed CS term has already been studied based on the following moose \cite{Hill:2004uc}:
\be
\begin{tabular}{c}
\xymatrix@R=.4pc@C=1.4pc{\mathrm{Global:} & SU(N_f)_L &&  SU(N_f)_R \\
& *=<20pt>[o][F]{} \doublerightxyarrow^{\mbox{\raisebox{1.5ex}{$\Sigma$}}} && *=<20pt>[o][F]{} \\ \mathrm{Gauged:} &SU(N_f)_L &&  SU(N_f)_R&}
\end{tabular}
\ee
where the gauged $SU(N_f)_i$ correspond to arbitary left- and right-handed currents.  Now  with inspiration from AdS/CFT, we want to understand the CS term in the $G/H$ moose, where the natural variable is $\xi$, the ``square root'' of $\Sigma$:
\be
\begin{tabular}{c}
\xymatrix@R=.4pc@C=1.4pc{\mathrm{Global:} & SU(N_f)_L \times  SU(N_f)_R && SU(N_f)_L \times  SU(N_f)_R \\
& *=<20pt>[o][F]{} \doublerightxyarrow^{\mbox{\raisebox{1.5ex}{$\xi$}}} && *=<20pt>[o][F]{} \\ \mathrm{Gauged:} &SU(N_f)_L \times  SU(N_f)_R &&SU(N_f)_V&}
\end{tabular}
\ee
In deriving CCWZ from the $G/H$ moose, we saw that an essential step in constructing the Goldstone kinetic terms was integrating out the $H_\mu$ field (here, the $SU(N_f)_V$ gauge bosons), and we expect that this will also be an essential ingredient in constructing the moose-ified WZW term.

\acknowledgments{We would like to thank Itay Yavin and Nima Arkani-Hamed for numerous discussions and many important insights.  We gratefully acknowledge conversations with Spencer Chang, Howard Georgi, and Martin Schmaltz.  This work was supported by an NSF graduate research fellowship.}

\appendix

\section{Warped Dimensions and Nonlinear Sigma Models}
\label{sec:warp}

In this appendix, we review how to identify the $A_5$ component of a bulk gauge field with a nonlinear sigma model link field $\xi$.  We derive the effective lagrangian for $\xi$ in any theory where the extra dimension is an interval and the 4D slices are Minkowski space.

We first note that this construction is unique to extra dimensions with boundaries.  If we had a compactified flat extra dimension with circumference $R$ and bulk gauge fields, there would be two massless states:  a spin-1 zero mode ($A^{(0)}_\mu$) and spin-0 zero mode ($A^{(0)}_5$).  However, with the minimal gauge kinetic lagrangian, there is zero amplitude for $A^{(0)}_5 A^{(0)}_5$ scattering, so the effective pion decay constant for $A^{(0)}_5$ is $f_\pi = \infty$.  We can see this in two ways.  First, by KK momentum conservation, there is no
\beq{threepoint}
A^{(0)}_5 A^{(0)}_5 \partial^\mu A^{(n)}_\mu
\eeq
vertex, so when we integrate out the heavy gauge modes, we do not generate a
\beq{fourpoint}
(A^{(0)}_5 \partial_\mu A^{(0)}_5)^2
\eeq
four-point interaction.  Second, we can imagine deconstructing the compactified dimension with $N$ link fields.  Each link $U_i$ has decay constant $f$, and the product of the link fields $U = U_1 U_2 \cdots U_N$ plays the role of the $A_5$ zero mode.  It is easy to show that the effective decay constant $f_\pi$ for $U$ is $f_\pi^2 = N f^2$.  In terms of the 5D gauge coupling $g_5$ and the lattice spacing $a = R/N$,
\be
f^2 = \frac{1}{g_5^2 a}, \qquad f_\pi^2 = Nf^2 = \frac{N}{g_5^2 a} = \frac{1}{g_4^2 a^2},
\ee
where $g_4 = g_5/\sqrt{R}$ is the effective 4D gauge coupling.  For fixed lattice spacing, $f_\pi$ is finite, but as we go to the continuum theory $a \rightarrow 0$, the pion decay constant $f_\pi$ goes to infinity.  In other words, with a homogeneous compact extra dimension, the $A_5$ zero mode is simply a free scalar field, and cannot be interpreted as a nonlinear sigma model.

When the extra dimension is an interval with fixed boundaries (or a compact space with defects \cite{Arkani-Hamed:2001nc}), there is no KK momentum conservation, and we can generate an effective $A^{(0)}_5$ four point interaction.  In flat space, we can easily estimate $f_\pi$.  The coefficient of the vertex in \eq{threepoint} is $g_4$ because it arises from gauge interactions.  For an interval of length $R$, the mass of the first KK mode is $1/R$, so integrating out the massive mode, the vertex in \eq{fourpoint} has coefficient $g_4^2 R^2$.  In terms of the pion decay constant,
\beq{flatfpi}
\frac{1}{f_\pi^2}(A^{(0)}_5 \partial_\mu A^{(0)}_5)^2, \qquad f_\pi = \frac{1}{g_4 R}.
\ee
As expected, $f_\pi$ is proportional to the only mass scale in the theory $1/R$, and goes to infinity as the gauge coupling goes to zero.

Of course, this argument assumed that such an $A^{(0)}_5$ zero mode existed, and this depends on the boundary conditions on the interval.  If $A_\mu$ has Neumann boundary conditions at either boundary, then we can simply go to $A_5 = 0$ gauge, and there is no candidate for the nonlinear sigma field.   The only way that there can be uneaten (psuedo-)Goldstone bosons is if some components of $A_\mu$ have Dirichlet boundary condition on both branes \cite{Contino:2003ve}, and looking at \eq{duality}, the $A_5$ zero modes in $G / (F \cup H)$ are indeed the uneaten Goldstones.

In an arbitrary warped space, we can figure out the pion decay constant for the Goldstones most simply by deconstructing the fifth dimension.  Consider a metric for an extra dimensional interval with 4D Minkowski slices:
\be
ds^2 = e^{-2\sigma(y)}\eta_{\mu\nu}dx^\mu dx^\nu - dy^2,
\ee  
where $0 < y < R$, and $\sigma(y)$ is an arbitrary function.  Assuming no boundary kinetic terms, the action for bulk $G$ gauge bosons is:
\be
S = -\frac{1}{2}\int \! d^4 x \, dy \,  \frac{1}{g_5^2} \tr \left( F_{\mu\nu}^2 -2 e^{-2\sigma} F^2_{\mu 5} \right).
\ee
We can deconstruct this action with $N+1$ sites, $N$ links, and equal lattice spacing $a$ \cite{Randall:2002qr}:
\be
\begin{tabular}{c}
\xymatrix@R=.4pc@C=1.4pc{A^0_\mu && A^1_\mu &&  && A^{N-1}_\mu && A^N_\mu \\ *=<20pt>[o][F]{} \doublerightxyarrow^{\mbox{\raisebox{1.5ex}{$U_1$}}} && *=<20pt>[o][F]{} \doublerightxyarrow^{\mbox{\raisebox{1.5ex}{$U_2$}}} && *=<20pt>[o]{\cdots} \doublerightxyarrow^{\mbox{\raisebox{1.5ex}{$U_{N-1}$}}} && *=<20pt>[o][F]{} \doublerightxyarrow^{\mbox{\raisebox{1.5ex}{$U_N$}}} && *=<20pt>[o][F]{}
}
\end{tabular}
\vspace{.15in}
\ee
The action for this system is
\beq{deconAdSaction}
S = \int \! d^4 x \left(-\frac{1}{2}\sum_{i=0}^N \frac{a}{g_5^2} \tr (F^{i}_{\mu\nu})^2 + \sum_{i=1}^N \frac{e^{-2\sigma_i}}{g_5^2 a} \tr |D_\mu U_i |^2 \right),
\ee
where $\sigma_i = \sigma(a i)$, and the link field covariant derivative is
\be
D_\mu U_i = \partial_\mu U_i + i A^{i-1}_\mu U_i - i U_i A^{i}_\mu.
\ee
At this point, we have not specified the boundary conditions on the end sites.  For simplicity, we take $A^0_\mu = 0$,  which corresponds to not gauging any subgroup of the $G$ global symmetry in the dual theory.  Similarly, we take $A^N_\mu \in H$ to account for the spontaneous breaking of $G$ to $H$ in the dual theory.   We can go to a unitary gauge where $U_i = 1$ for $1\le i \le N-1$ and $U_N \equiv \xi \in G/H$.  The spectrum consists of the massless Goldstones in $\xi$, and a tower of massive spin-1 fields.    

It is straightforward to integrate out these heavy spin-1 fields to arrive at an effective lagrangian for $\xi$.  Going to canonical normalization for the gauge fields and ignoring the gauge kinetic terms, \eq{deconAdSaction} becomes with our gauge choice:
\beq{unitarygaugeAdSaction}
\mathcal{L} = m_1^2 \tr (A^1_\mu)^2 + \sum_{i=2}^{N-1} m_i^2 \tr  \left(A^{i-1}_\mu - A^{i}_\mu \right)^2 + m_N^2 \tr \left| \frac{\sqrt{a}}{g_5}\partial_\mu \xi +i A^{N-1}_\mu \xi - i \xi A^{N}_\mu \right|^2,
\ee
where $m_i = e^{-\sigma_i}/a$.  Integrating out the massive $A_i$ fields:
\beq{prexikinetic} 
\mathcal{L} = f_{\rm eff}^2 \tr p^\mu p^\dagger_\mu,
\eeq
where $p_\mu$ is given by \eq{vpdecomp} with $F_\mu = 0$, and
\beq{discretefpi}
\frac{1}{f_{\rm eff}^2} = g_5^2 a \sum_{i=1}^N e^{2 \sigma_i}.
\ee
At we saw in section \ref{sec:intro}, this is precisely the CCWZ kinetic term for a $G/H$ nonlinear sigma field.

Going to the continuum limit of \eq{discretefpi}, the pion decay constant for $\xi$ is
\beq{generalfequation}
\frac{1}{f_{\rm eff}^2} = g_4^2 R \int \! dy\, e^{2 \sigma(y)}.
\ee
In the case of a flat extra dimension, $f_{\rm eff} = 1/ g_4 R$ as we guessed in \eq{flatfpi}.  In AdS space, the warp function is $\sigma(y)= ky$, and in the limit $kR \gg 1$,
\be
f_{\rm eff} = \frac{e^{-kR} k \sqrt{2}}{g_4 \sqrt{kR}}.
\ee
This result is parametrically what we expect from AdS/CFT.  It is customary to define the AdS perturbative expansion parameter $g_\rho = g_4 \sqrt{kR}$, which can be holographically identified as $g_\rho \sim 4\pi/\sqrt{N}$ in a large $N$ CFT \cite{Arkani-Hamed:2000ds}.  The scale $k e^{-kR}$ can be roughly identified with the confinement scale $\Lambda_{QCD}$.  So we see that in AdS space
\be
f_{\rm eff} \sim \frac{\Lambda_{QCD}}{4\pi}\sqrt{N},
\ee
giving the expected $\sqrt{N}$ scaling of $f_{\rm eff}$ from the dual CFT description \cite{'tHooft:1973jz}.

\section{Boundary Condition Breaking vs. Linear Sigma Models}
\label{sec:bc}

In section \ref{sec:intro}, we were implicitly using Dirichlet boundary conditions to reduce the gauge symmetry on the UV and IR branes when we deconstructed AdS$_5$.  One might worry that this construction is not entirely general because it is also possible to reduce the symmetries by linear sigma fields on the branes (or even by brane-localized technicolor).  In AdS model building, the vev of the linear sigma field is often a free parameter that is adjusted to match phenomenology.  Boundary condition breaking is the limit that the vev of the linear sigma field goes to infinity, and it may seem like we are neglecting an additional free parameter by always going to this limit.  

In the case that the vev is much smaller than the scale of the IR brane, then we might as well not even talk about AdS space, because the low energy degrees of freedom (including the radial modes of the linear sigma field) are well separated from the AdS KK excitations.  In the case that the vev is comparable to or larger than the IR brane scale, then the details of how the symmetries are broken are irrelevant for low energy physics, so all we care about is the resulting nonlinear sigma field on the IR brane after symmetry breaking.  

We will see that a nonlinear sigma field on the IR brane is equivalent to a different AdS space with boundary condition breaking on the IR brane but with a slightly distorted warp factor.  (An identical story holds for UV brane symmetry breaking.)  To make this discussion more concrete, we will imagine that $G/H$ is a symmetric space.  Let $T^a$ be the generators of $H$ and $X^a$ be the generators of $G/H$.  A symmetric space has a $\mathbf{Z}_2$ automorphism under which
\beq{z2auto}
T^a \rightarrow T^a, \qquad X^a \rightarrow -X^a.
\ee
We can then construct the object $\Sigma = \xi \tilde{\xi}^\dagger = \xi^2$, where tildes indicate the image of a field under \eq{z2auto}.  $\Sigma$ has the nice property that it transforms linearly under $G$
\be
\Sigma \rightarrow g \Sigma \tilde{g}^\dagger,
\ee
so we can imagine $\Sigma$ arising from a linear sigma field $\Phi \rightarrow g \Phi \tilde{g}^\dagger$ that gets a vev $\langle \Phi \rangle = \openone$.  If an $F$ subgroup of $G$ is gauged, the kinetic term for $\Sigma$ is
\beq{sigmakinetic}
\mathcal{L}_\Sigma = \frac{f^2}{4} \tr |D_\mu \Sigma|^2, \qquad D_\mu \Sigma = \partial_\mu \Sigma + i F_\mu \Sigma - i \Sigma \tilde{F}_\mu.
\ee
Using the $\mathbf{Z}_2$ automorphism on \eq{vpdecomp} it is easy to show that 
\be
\frac{f^2}{4} \tr |D_\mu \Sigma|^2 = f^2 \tr p^\mu p^\dagger_\mu,
\ee
as in \eq{ccwzlagrange}.  In other words, the following moose diagrams have identical low energy physics:
\be
\begin{tabular}{c}
\xymatrix@R=.4pc@C=1.4pc{G_{\rm global} \\
 *=<20pt>[o][F]{} \ar@`{{?+(20,20),?+(20,-20)}}@{-}[] |{\SelectTips{eu}{}\object@{>}}^{\mbox{\raisebox{0ex}{$\, \Sigma$}}}  }
\end{tabular}
\qquad
\begin{tabular}{c}
\xymatrix@R=.4pc@C=1.4pc{G_{\rm global} && H_{\rm gauged} \\
*=<20pt>[o][F]{} \doublerightxyarrow^{\mbox{\raisebox{1.5ex}{$\xi$}}} && *=<20pt>[o][F]{}}
\end{tabular}
\vspace{.15in}
\ee
where it is understood that $\Sigma$ transforms as $\Sigma \rightarrow g \Sigma \tilde{g}^\dagger$.

If we insert the nonlinear sigma field on the IR brane, the $G$ global symmetry is identified with the bulk $G$ gauge symmetry.  Deconstructing the AdS space, we have two theories with identical low energy physics:
\beq{twomooses}
\begin{tabular}{c}
\xymatrix@R=.4pc@C=1.4pc{ & G_{\rm gauged} & G_{\rm gauged} & G_{\rm gauged} \\ \cdots \rightxyarrow &  *=<20pt>[o][F]{} \rightxyarrow &  *=<20pt>[o][F]{}\rightxyarrow &
 *=<20pt>[o][F]{} \ar@`{{**{} ?+(20,20),?+(20,-20)}}@{-}[] |{\SelectTips{eu}{}\object@{>}}^{\mbox{\raisebox{0ex}{$\, \Sigma$}}} \\
 \\
   & G_{\rm gauged} & G_{\rm gauged} & G_{\rm gauged} && H_{\rm gauged} \\
 \cdots \rightxyarrow &  *=<20pt>[o][F]{} \rightxyarrow &  *=<20pt>[o][F]{}\rightxyarrow & *=<20pt>[o][F]{} \doublerightxyarrow^{\mbox{\raisebox{1.5ex}{$\xi$}}} && *=<20pt>[o][F]{}
 }
\end{tabular}
\bigskip
\ee
The first moose is just the deconstructed version of nonlinear sigma field living on the IR brane.  The second moose can now be interpreted as a deconstructed AdS space where the IR brane is slightly shifted and boundary conditions reduce the gauge symmetry to $H$ on the IR brane.   In order to account for the fact that the decay constant on the $\xi$ link might not track the standard warp factor when we reconstruct the AdS space, we would have to allow warp factor to be slightly perturbed near the IR brane.  

In fact, as we have already seen in \eq{littlesimilar}, we could put the $N$ site moose of \eq{nsitemoose} on the IR brane because this moose also describes a $G/H$ nonlinear sigma model at low energies.  The point is that it hardly matters how one breaks $G \rightarrow H$ on the IR brane because all the relevant information for how it is broken can be absorbed into a redefinition of the warp factor near the IR brane.  Similarly, it is often computationally simpler to work with an IR brane nonlinear sigma field instead of boundary condition breaking, and if we look at \eq{discretefpi} in appendix \ref{sec:warp}, we see that the decay constant for $\Sigma$ (which is the same as the decay constant for $\xi$) can be taking to infinity but the effective $G/H$ pion decay constant will still be finite.

Another interesting question with respect to linear sigma models and boundary condition breaking is what fields are responsible for unitarizing Goldstone-Goldstone scattering in AdS space.  This is relevant both in Higgsless theories \cite{Csaki:2003dt}, where the Goldstones are the longitudinal modes of the $W$ and $Z$ bosons, and in composite Higgs models \cite{Contino:2003ve,Agashe:2004rs} where some of the Goldstones comprise the Higgs boson.  In the context of the littlest Higgs, there should be fields in the AdS construction of \cite{Thaler:2005en} that restore perturbative unitarity of Goldstone scattering \cite{Chang:2003vs}.  If $G$ is broken to $H$ on the IR brane through a linear sigma model \emph{and} the vev of the linear sigma field is much smaller than the IR brane scale, then unitarity is restored by the radial modes of the linear sigma field, in analogy with the ordinary Higgs mechanism.  

When the linear sigma vev is comparable to the IR brane scale, however, the IR brane Goldstones mix with the bulk gauge fields, and it is no longer accurate to think of the physical Goldstones as being entirely localized on the IR brane.  In that case, one expects that unitarity is restored by some combination of the radial modes of the linear sigma field and the KK gauge bosons that have the same quantum numbers as the radial modes.  The radial modes fill an adjoint and a singlet of $H$, and there are certainly KK gauge bosons that transform as adjoint of $H$, namely those with Neumann boundary conditions on the IR brane.  In the limit that the linear sigma vev goes to infinity, the KK gauge bosons with Dirichlet boundary conditions on the IR brane are singlets under $H$, and we also expect that there will be other singlet states corresponding to broad resonances of the CFT (``QCD string'' states) \cite{Arkani-Hamed:2000ds}.   The point which is made obvious by the two-site deconstruction is that even when the linear sigma vev goes to infinity, there is still a state in the theory that has the quantum numbers to potentially address Goldstone-scattering unitarity, namely the $H_\mu$ field.  Whether unitarity is actually restored for arbitrary boundary conditions of course requires a more complete analysis \cite{Csaki:2003dt}.

\section{Fermions and the Moose-ified Littlest Higgs}
\label{sec:fermions}

To construct a realistic littlest Higgs theory, we have to include the fermion sector of the standard model.  In this appendix, we sketch how to incorporate the top quark into the $N_f = 5$ QCD UV completion of the littlest Higgs from \eq{littlehiggsmoose}:
\beq{littlehiggsmoose2}
\begin{tabular}{c}
\xymatrix@R=.4pc@C=1.4pc{\mathrm{Global:} & SU(5)_L && SU(5)_R \\
& *=<20pt>[o][F]{} \rightxyarrow^{\mbox{\raisebox{1.5ex}{$\psi$}}} & *=<12pt>[o][F=]{} \rightxyarrow^{\mbox{\raisebox{1.5ex}{$\psi^c$}}} & *=<20pt>[o][F]{} \\ \mathrm{Gauged:} &SU(2)^2 \times U(1)_Y &SU(N_c)&SO(5)}
\end{tabular}
\ee
When $SU(N_c)$ confines, $\langle \psi \psi^c \rangle$ condenses to become the link field $\xi$.  We will see that while it is certainly possible to include standard model fermions in the moose-ified littlest Higgs, we will run into many of the same challenges from technicolor in trying to generate a large enough top Yukawa coupling.  Also, we will ignore any possible gauge anomalies, though for generic fermion content, there will be anomalies associated with the $SU(2)^2 \times U(1)_Y$ gauge group.  

In the original language of global $SU(5)/SO(5)$ breaking \cite{Arkani-Hamed:2002qy}, the relevant Goldstones were packaged into the field $\Sigma = \xi^2 \Sigma_0$, where
\be
\Sigma_0 \equiv \left(
\begin{array}{ccc} &  & \openone \\ & 1 &  \\ \openone &  &
\end{array}\right).
\ee
$\Sigma$ transforms as $\Sigma \rightarrow V \Sigma V^T$, where $V \in SU(5)$.  The Higgs doublet comprises four of the Goldstones in $\xi$.   We will use the gauge sector and fermion content of \cite{Katz:2003sn}.  In that case, there is an $SU(2)^2 \times U(1)_A$ subgroup of $SU(5)$ generated by
\be
Q_1^a =\left(\begin{array}{cc}\sigma^{a}/2 & \quad \\ \quad&\quad \end{array}\right), \qquad Q_2^a = \left(\begin{array}{cc} \quad&\quad \\ \quad &-\sigma^{*a}/2\end{array}\right), \qquad A = \diag(1,1,0,-1,-1)/2.
\ee
Both $SU(2)$ subgroups are gauged, and the breaking of $SU(5) \rightarrow SO(5)$ Higgses $SU(2)^2$ to the diagonal $SU(2)_{EW}$.  There is a separate $U(1)_B$ group, and hypercharge is generated by $Y = A+B$.

The fermion lagrangian of the littlest Higgs can be written compactly as
\beq{fermionsector}
\mathcal{L} = f \left(  \lambda_1 Q \Sigma^\dagger Q^c  + \lambda_2 \tilde{Q} \Sigma_0 Q^c  + \lambda_3 Q \Sigma_0 \tilde{Q}^c+ h.c. \right),
\ee
where $f$ is the decay constant of $\xi$, and the fermions transform as:
\be
\begin{tabular}{l|ccc}
 & $SU(3)_C$ & $SU(5)$ & $U(1)_B$ \\
 \hline
$Q$ & $\mathbf{3}$ & $\mathbf{5}$& $+2/3$\\
$Q^c$ & $\mathbf{\bar{3}}$ & $\mathbf{5}$& $-2/3$\\
$\tilde{Q}$ & $\mathbf{3}$ &  $\mathbf{\bar{5}}$& $+2/3$\\
$\tilde{Q}^c$ & $\mathbf{\bar{3}}$ &  $\mathbf{\bar{5}}$ & $-2/3$
\end{tabular}
\ee
where $SU(3)_C$ is the color gauge group.  As written, the lagrangian in \eq{fermionsector} does not break any of the $SU(5)$ symmetries that protect the Higgs.  However, $SU(5)$ is explicitly broken because $\tilde{Q}$ and $\tilde{Q}^c$ are incomplete $SU(5)$ multiplets:
\be
Q = \left(\begin{array}{c}p \\ \tilde{t} \\ q \end{array}\right), \qquad Q^c = \left(\begin{array}{c}\tilde{q}^c \\t^c \\ p^c \end{array}\right), \qquad \tilde{Q} = \left(\begin{array}{c}0 \\0 \\ \tilde{q} \end{array}\right), \qquad \tilde{Q}^c = \left(\begin{array}{c}0 \\ \tilde{t}^c \\0\end{array}\right).
\ee
In order for the Higgs to acquire a radiative potential from the top sector, each $\lambda_i$ must be non-zero, and therefore \eq{fermionsector} manifestly exhibits collective breaking.  (See \cite{Katz:2003sn,Thaler:2005en} for a more detailed discussion of the structure and purpose of this fermion content.)  Couplings to other standard model fermions can be included by adding explicit $SU(5)$ violating couplings to $\Sigma$.

To simplify the discussion a bit, we will work in a slightly different $SU(5)$ basis where $\Sigma_0 = \openone$.  The unbroken $SO(5)$ generators $T_a$ and the broken $SU(5)/SO(5)$ generators $X_a$ satisfy
\be
T_a   = - T_a^T, \qquad X_a = X_a^T.
\ee
We see readily that $SU(5)/SO(5)$ is a symmetric space where the $\mathbf{Z}_2$ automorphism is $Q_a \rightarrow - Q_a^T$.  Note that $\xi = e^{i \pi_a X_a}$, so $\xi  = \xi^T$.  For this reason, is convenient to make explicit the implied transposes in \eq{fermionsector}:
\beq{fermionsector2}
\mathcal{L} = f \left(  \lambda_1 Q^T \xi^* \xi^\dagger Q^c  + \lambda_2 \tilde{Q}^T Q^c  + \lambda_3 Q^T \tilde{Q}^c+ h.c. \right).
\ee
We are already anticipating that $\xi$ will transforms as $\xi \rightarrow L \xi R^\dagger$ under $SU(5)_L \times SU(5)_R$ as suggested by \eq{littlehiggsmoose2}.  Note that $\xi \xi^T$ is indeed invariant under $SO(5)_R$, which is necessary for the gauge invariance of the QCD moose.

There are three simple ways to embed the fermions from \eq{fermionsector2} into the QCD moose.  The first is to simply identify the original $SU(5)$ with $SU(5)_L$, in which case the $\lambda_1$ interaction would arise from the dimension nine operator
\be
\frac{1}{M_{\rm NP}^5}Q^T (\psi \psi^c)^* (\psi \psi^c)^\dagger Q^c,
\ee
where we are imagining that there is some new physics at $M_{\rm NP}$ that generates this interaction.  This is a highly irrelevant operator, and unless we postulate some large anomalous dimension for the condensate $\langle \psi \psi^c \rangle$ \cite{Holdom:1981rm}, it seems highly unlikely to ever end up with an $\mathcal{O}(1)$ $\lambda_1$ if $M_{\rm NP}$ is much different from $\Lambda_{\rm QCD}$.  

Another way to include fermions is to take a cue from the AdS$_5$ construction of \cite{Thaler:2005en}.  In that case, $Q$ and $Q^c$ are bulk fermions, so in the moose of \eq{littlehiggsmoose2}, we should have $Q$ and $Q^c$ fermions charged under $SU(5)_L$ and $\chi$ and $\chi^c$ fermions charged under $SU(5)_R$ with the same $SU(3)_C$ and $U(1)_B$ charges as their counterparts.  The generalization of \eq{fermionsector2} is
\be
\mathcal{L} = f \left( \lambda_{1a} Q^T \xi^* \chi^c +  \lambda_{1b} \chi^T \xi^\dagger Q^c + \lambda_{1c} \chi^T \chi^c + \lambda_2 \tilde{Q}^T Q^c  + \lambda_3 Q^T \tilde{Q}^c + h.c. \right),
\ee
which is invariant under $SU(5)_L \times SU(5)_R$ when $SU(5)_R$ is restricted to $SO(5)_R$.  Integrating out $\chi$ and $\chi^c$, we see that $\lambda_1 = \lambda_{1a} \lambda_{1b}/\lambda_{1c}$.  In terms of the fermions $\psi$ and $\psi^c$, we could generate the appropriate four fermion operators through an extended technicolor-like mechanism \cite{Eichten:1979ah,Dimopoulos:1979es}:
\be
\frac{(4\pi)^2}{M_{\rm ETC}^2}Q^T (\psi \psi^c)^* \chi^c, \qquad \frac{(4\pi)^2}{M_{\rm ETC}^2} \chi^T (\psi \psi^c)^\dagger Q^c, \qquad m_\chi \chi^T \chi^c,
\ee
where we are assuming that all gauge couplings are large ($\sim 4 \pi$).  If the condensate $\langle \psi \psi^c \rangle$ takes its NDA value $4\pi f^3$, then parametrically
\be
\lambda_1 \sim \frac{4\pi \Lambda_{\rm QCD}^5}{M_{\rm ETC}^4 m_\chi}, \qquad  \Lambda_{\rm QCD} \sim 4\pi f.
\ee
In order to have $\lambda_1 \sim \mathcal{O}(1)$, the ETC scale must be quite low or the $\langle \psi \psi^c \rangle$ must have a large anomalous dimension.  Also, to have a realistic top Yukawa coupling and a light $t'$ fermion partner, all of the $\lambda_i$ values must be roughly the same, and in this scenario, there is no explanation for why the $\tilde{Q} Q^c$ and $Q \tilde{Q}^c$ mass terms will be roughly the same size as the $Q \Sigma^\dagger Q^c$ interaction.

Finally, we can rewrite \eq{fermionsector2} equivalently as
\be
\mathcal{L} = f \left(  \lambda_1 (\xi^\dagger Q)^T (\xi^\dagger Q^c)  + \lambda_2 \tilde{Q}^T \xi (\xi^\dagger Q^c)  + \lambda_3 (\xi^\dagger Q)^T \xi^T \tilde{Q}^c+ h.c. \right).
\ee
This suggests that we could identify $\xi^\dagger Q \equiv Q'$ and $\xi^\dagger Q^c \equiv Q'^c$ as fundamentals of $SU(5)_R$:
\be
\mathcal{L} = f \left(  \lambda_1 Q'^T Q'^c  + \lambda_2 \tilde{Q}^T \xi Q'^c  + \lambda_3 Q'^T \xi^T \tilde{Q}^c+ h.c. \right).
\ee
Though it might appear that $Q'$ and $Q'^c$ have lost information about their $SU(2)^2$ charges, recall that when we integrate out the $SO(5)$ gauge bosons, couplings to the gauged subgroup of $SU(5)_L$ are induced as in \eq{fermionvmucoupling}.  We can now use an ETC-like mechanism to generate the operators
\be
\frac{(4\pi)^2}{M_{\rm ETC}^2}\tilde{Q}^T (\psi \psi^c) Q'^c, \qquad \frac{(4\pi)^2}{M_{\rm ETC}^2}Q'^T (\psi \psi^c)^T \tilde{Q}^c.
\ee
Again, without a low ETC scale or a large $\langle \psi \psi^c \rangle$ condensate vev, it will be difficult to generate a large enough top Yukawa coupling, and there is no reason why the $\lambda_i$ values should be similar. 

Of course, apart from the $\lambda_i$ degeneracy problem, these difficulties in generating realistic Yukawa couplings are identical to issues in technicolor.  To the extent that little Higgs theories predict perturbative physics well above the the electroweak scale, the moose-ified littlest Higgs is an improvement over technicolor in that it has a realistic chance to address precision electroweak constraints.  The lack of a GIM mechanism \cite{Glashow:1970gm} in extended technicolor means that the implementation of fermions presented in this section will most likely generate large sources of flavor violation unless we are able to raise the ETC scale via some walking mechanism.    But in terms of the particle content of the littlest Higgs (if not the actual numeric spectrum of the theory), we have seen that it is fully consistent to embed standard model fermions in an $N_f = 5$ QCD UV completion of the littlest Higgs.


\begin{thebibliography}{99}


%\cite{Gross:1973id}
\bibitem{Gross:1973id}
D.~J.~Gross and F.~Wilczek,
\emph{Ultraviolet Behavior Of Non-Abelian Gauge Theories,}
Phys.\ Rev.\ Lett.\  {\bf 30}, 1343 (1973).
%%CITATION = PRLTA,30,1343;%%

%\cite{Politzer:1973fx}
\bibitem{Politzer:1973fx}
H.~D.~Politzer,
\emph{Reliable Perturbative Results For Strong Interactions?,}
Phys.\ Rev.\ Lett.\  {\bf 30}, 1346 (1973).
%%CITATION = PRLTA,30,1346;%%

%\cite{Arkani-Hamed:2001nc}
\bibitem{Arkani-Hamed:2001nc}
N.~Arkani-Hamed, A.~G.~Cohen and H.~Georgi,
\emph{Electroweak symmetry breaking from dimensional deconstruction,}
Phys.\ Lett.\ B {\bf 513}, 232 (2001)
[arXiv:hep-ph/0105239].
%%CITATION = HEP-PH 0105239;%%

%\cite{Arkani-Hamed:2002pa}
\bibitem{Arkani-Hamed:2002pa}
N.~Arkani-Hamed, A.~G.~Cohen, T.~Gregoire and J.~G.~Wacker,
\emph{Phenomenology of electroweak symmetry breaking from theory space,}
JHEP {\bf 0208}, 020 (2002)
[arXiv:hep-ph/0202089].
%%CITATION = HEP-PH 0202089;%%


%\cite{Arkani-Hamed:2002qy}
\bibitem{Arkani-Hamed:2002qy}
N.~Arkani-Hamed, A.~G.~Cohen, E.~Katz and A.~E.~Nelson,
\emph{The littlest Higgs,}
JHEP {\bf 0207}, 034 (2002)
[arXiv:hep-ph/0206021].
%%CITATION = HEP-PH 0206021;%%

%\cite{Arkani-Hamed:2002qx}
\bibitem{Arkani-Hamed:2002qx}
N.~Arkani-Hamed, A.~G.~Cohen, E.~Katz, A.~E.~Nelson, T.~Gregoire and J.~G.~Wacker,
\emph{The minimal moose for a little Higgs,}
JHEP {\bf 0208}, 021 (2002)
[arXiv:hep-ph/0206020].
%%CITATION = HEP-PH 0206020;%%

%\cite{Gregoire:2002ra}
\bibitem{Gregoire:2002ra}
T.~Gregoire and J.~G.~Wacker,
\emph{Mooses, topology and Higgs,}
JHEP {\bf 0208}, 019 (2002)
[arXiv:hep-ph/0206023].
%%CITATION = HEP-PH 0206023;%%

%\cite{Low:2002ws}
\bibitem{Low:2002ws}
I.~Low, W.~Skiba and D.~Smith,
\emph{Little Higgses from an antisymmetric condensate,}
Phys.\ Rev.\ D {\bf 66}, 072001 (2002)
[arXiv:hep-ph/0207243].
%%CITATION = HEP-PH 0207243;%%

%\cite{Kaplan:2003uc}
\bibitem{Kaplan:2003uc}
D.~E.~Kaplan and M.~Schmaltz,
\emph{The little Higgs from a simple group,}
JHEP {\bf 0310}, 039 (2003)
[arXiv:hep-ph/0302049].
%%CITATION = HEP-PH 0302049;%%

%\cite{Skiba:2003yf}
\bibitem{Skiba:2003yf}
W.~Skiba and J.~Terning,
\emph{A simple model of two little Higgses,}
Phys.\ Rev.\ D {\bf 68}, 075001 (2003)
[arXiv:hep-ph/0305302].
%%CITATION = HEP-PH 0305302;%%

%\cite{Cheng:2003ju}
\bibitem{Cheng:2003ju}
H.~C.~Cheng and I.~Low,
\emph{TeV symmetry and the little hierarchy problem,}
JHEP {\bf 0309}, 051 (2003)
[arXiv:hep-ph/0308199].
%%CITATION = HEP-PH 0308199;%%

%\cite{Cheng:2004yc}
\bibitem{Cheng:2004yc}
H.~C.~Cheng and I.~Low,
\emph{Little hierarchy, little Higgses, and a little symmetry,}
JHEP {\bf 0408}, 061 (2004)
[arXiv:hep-ph/0405243].
%%CITATION = HEP-PH 0405243;%%


%\cite{Glashow:1970gm}
\bibitem{Glashow:1970gm}
S.~L.~Glashow, J.~Iliopoulos and L.~Maiani,
\emph{Weak Interactions With Lepton-Hadron Symmetry,}
Phys.\ Rev.\ D {\bf 2}, 1285 (1970).
%%CITATION = PHRVA,D2,1285;%%

%\cite{Katz:2003sn}
\bibitem{Katz:2003sn}
E.~Katz, J.~y.~Lee, A.~E.~Nelson and D.~G.~E.~Walker,
\emph{A composite little Higgs model,}
arXiv:hep-ph/0312287.
%%CITATION = HEP-PH 0312287;%%



%\cite{Weinberg:1975gm}
\bibitem{Weinberg:1975gm}
S.~Weinberg,
\emph{Implications Of Dynamical Symmetry Breaking,}
Phys.\ Rev.\ D {\bf 13}, 974 (1976).
%%CITATION = PHRVA,D13,974;%%

%\cite{Susskind:1978ms}
\bibitem{Susskind:1978ms}
L.~Susskind,
\emph{Dynamics Of Spontaneous Symmetry Breaking In The Weinberg-Salam Theory,}
Phys.\ Rev.\ D {\bf 20}, 2619 (1979).
%%CITATION = PHRVA,D20,2619;%%



%\cite{Coleman:1969sm}
\bibitem{Coleman:1969sm}
S.~R.~Coleman, J.~Wess and B.~Zumino,
\emph{Structure Of Phenomenological lagrangians. 1,}
Phys.\ Rev.\  {\bf 177}, 2239 (1969).
%%CITATION = PHRVA,177,2239;%%

%\cite{Callan:1969sn}
\bibitem{Callan:1969sn}
C.~G.~Callan, S.~R.~Coleman, J.~Wess and B.~Zumino,
\emph{Structure Of Phenomenological lagrangians. 2,}
Phys.\ Rev.\  {\bf 177}, 2247 (1969).
%%CITATION = PHRVA,177,2247;%%



%\cite{Bando:1987br}
\bibitem{Bando:1987br}
M.~Bando, T.~Kugo and K.~Yamawaki,
\emph{Nonlinear Realization And Hidden Local Symmetries,}
Phys.\ Rept.\  {\bf 164}, 217 (1988).
%%CITATION = PRPLC,164,217;%%


%\cite{Georgi:1989xy}
\bibitem{Georgi:1989xy}
H.~Georgi,
\emph{Vector Realization Of Chiral Symmetry,}
Nucl.\ Phys.\ B {\bf 331}, 311 (1990).
%%CITATION = NUPHA,B331,311;%%

%\cite{Contino:2003ve}
\bibitem{Contino:2003ve}
R.~Contino, Y.~Nomura and A.~Pomarol,
\emph{Higgs as a holographic pseudo-Goldstone boson,}
Nucl.\ Phys.\ B {\bf 671}, 148 (2003)
[arXiv:hep-ph/0306259].
%%CITATION = HEP-PH 0306259;%%


%\cite{Wess:1971yu}
\bibitem{Wess:1971yu}
J.~Wess and B.~Zumino,
\emph{Consequences Of Anomalous Ward Identities,}
Phys.\ Lett.\ B {\bf 37}, 95 (1971).
%%CITATION = PHLTA,B37,95;%%

%\cite{Witten:1983tw}
\bibitem{Witten:1983tw}
E.~Witten,
\emph{Global Aspects Of Current Algebra,}
Nucl.\ Phys.\ B {\bf 223}, 422 (1983).
%%CITATION = NUPHA,B223,422;%%


%\cite{Maldacena:1997re}
\bibitem{Maldacena:1997re}
J.~M.~Maldacena,
\emph{The large N limit of superconformal field theories and supergravity,}
Adv.\ Theor.\ Math.\ Phys.\  {\bf 2}, 231 (1998)
[Int.\ J.\ Theor.\ Phys.\  {\bf 38}, 1113 (1999)]
[arXiv:hep-th/9711200].
%%CITATION = HEP-TH 9711200;%%

%\cite{Gubser:1998bc}
\bibitem{Gubser:1998bc}
S.~S.~Gubser, I.~R.~Klebanov and A.~M.~Polyakov,
\emph{Gauge theory correlators from non-critical string theory,}
Phys.\ Lett.\ B {\bf 428}, 105 (1998)
[arXiv:hep-th/9802109].
%%CITATION = HEP-TH 9802109;%%

%\cite{Witten:1998qj}
\bibitem{Witten:1998qj}
E.~Witten,
\emph{Anti-de Sitter space and holography,}
Adv.\ Theor.\ Math.\ Phys.\  {\bf 2}, 253 (1998)
[arXiv:hep-th/9802150].
%%CITATION = HEP-TH 9802150;%%

%\cite{Arkani-Hamed:2000ds}
\bibitem{Arkani-Hamed:2000ds}
N.~Arkani-Hamed, M.~Porrati and L.~Randall,
\emph{Holography and phenomenology,}
JHEP {\bf 0108}, 017 (2001)
[arXiv:hep-th/0012148].
%%CITATION = HEP-TH 0012148;%%

%\cite{Rattazzi:2000hs}
\bibitem{Rattazzi:2000hs}
R.~Rattazzi and A.~Zaffaroni,
\emph{Comments on the holographic picture of the Randall-Sundrum model,}
JHEP {\bf 0104}, 021 (2001)
[arXiv:hep-th/0012248].
%%CITATION = HEP-TH 0012248;%%

%\cite{Randall:1999ee}
\bibitem{Randall:1999ee}
L.~Randall and R.~Sundrum,
\emph{A large mass hierarchy from a small extra dimension,}
Phys.\ Rev.\ Lett.\  {\bf 83}, 3370 (1999)
[arXiv:hep-ph/9905221].
%%CITATION = HEP-PH 9905221;%%

%\cite{Thaler:2005en}
\bibitem{Thaler:2005en}
J.~Thaler and I.~Yavin,
\emph{The Littlest Higgs in Anti-de Sitter Space,}
arXiv:hep-ph/0501036.
%%CITATION = HEP-PH 0501036;%%

%\cite{Arkani-Hamed:2001ca}
\bibitem{Arkani-Hamed:2001ca}
N.~Arkani-Hamed, A.~G.~Cohen and H.~Georgi,
\emph{(De)constructing dimensions,}
Phys.\ Rev.\ Lett.\  {\bf 86}, 4757 (2001)
[arXiv:hep-th/0104005].
%%CITATION = HEP-TH 0104005;%%


%\cite{Randall:2002qr}
\bibitem{Randall:2002qr}
L.~Randall, Y.~Shadmi and N.~Weiner,
\emph{Deconstruction and gauge theories in AdS(5),}
JHEP {\bf 0301}, 055 (2003)
[arXiv:hep-th/0208120].
%%CITATION = HEP-TH 0208120;%%


%\cite{Manohar:1983md}
\bibitem{Manohar:1983md}
A.~Manohar and H.~Georgi,
\emph{Chiral Quarks And The Nonrelativistic Quark Model,}
Nucl.\ Phys.\ B {\bf 234}, 189 (1984).
%%CITATION = NUPHA,B234,189;%%

%\cite{Georgi:1986kr}
\bibitem{Georgi:1986kr}
H.~Georgi and L.~Randall,
\emph{Flavor Conserving CP Violation In Invisible Axion Models,}
Nucl.\ Phys.\ B {\bf 276}, 241 (1986).
%%CITATION = NUPHA,B276,241;%%

%\cite{Bando:1984ej}
\bibitem{Bando:1984ej}
M.~Bando, T.~Kugo, S.~Uehara, K.~Yamawaki and T.~Yanagida,
\emph{Is Rho Meson A Dynamical Gauge Boson Of Hidden Local Symmetry?,}
Phys.\ Rev.\ Lett.\  {\bf 54}, 1215 (1985).
%%CITATION = PRLTA,54,1215;%%

%\cite{Harada:2003jx}
\bibitem{Harada:2003jx}
  M.~Harada and K.~Yamawaki,
  \emph{Hidden local symmetry at loop: A new perspective of composite gauge boson
  and chiral phase transition,}
  Phys.\ Rept.\  {\bf 381}, 1 (2003)
  [arXiv:hep-ph/0302103].
  %%CITATION = HEP-PH 0302103;%%

%\cite{Bando:1985rf}
\bibitem{Bando:1985rf}
  M.~Bando, T.~Kugo and K.~Yamawaki,
  \emph{On The Vector Mesons As Dynamical Gauge Bosons Of Hidden Local Symmetries,}
  Nucl.\ Phys.\ B {\bf 259}, 493 (1985).
  %%CITATION = NUPHA,B259,493;%%


%\cite{Bando:1987ym}
\bibitem{Bando:1987ym}
M.~Bando, T.~Fujiwara and K.~Yamawaki,
\emph{Generalized Hidden Local Symmetry And The A1 Meson,}
Prog.\ Theor.\ Phys.\  {\bf 79}, 1140 (1988).
%%CITATION = PTPKA,79,1140;%%

%\cite{Son:2003et}
\bibitem{Son:2003et}
D.~T.~Son and M.~A.~Stephanov,
\emph{QCD and dimensional deconstruction,}
Phys.\ Rev.\ D {\bf 69}, 065020 (2004)
[arXiv:hep-ph/0304182].
%%CITATION = HEP-PH 0304182;%%

%\cite{Piai:2004yb}
\bibitem{Piai:2004yb}
M.~Piai, A.~Pierce and J.~Wacker,
\emph{Composite vector mesons from QCD to the little Higgs,}
arXiv:hep-ph/0405242.
%%CITATION = HEP-PH 0405242;%%

%\cite{Harada:2003xa}
\bibitem{Harada:2003xa}
  M.~Harada, M.~Tanabashi and K.~Yamawaki,
  \emph{$\pi^+$ -- $\pi^0$ mass difference in the hidden local symmetry: A dynamical  origin
  of little Higgs,}
  Phys.\ Lett.\ B {\bf 568}, 103 (2003)
  [arXiv:hep-ph/0303193].
  %%CITATION = HEP-PH 0303193;%%



%\cite{Agashe:2004rs}
\bibitem{Agashe:2004rs}
K.~Agashe, R.~Contino and A.~Pomarol,
\emph{The minimal composite Higgs model,}
arXiv:hep-ph/0412089.
%%CITATION = HEP-PH 0412089;%%



%\cite{Eichten:1979ah}
\bibitem{Eichten:1979ah}
E.~Eichten and K.~D.~Lane,
\emph{Dynamical Breaking Of Weak Interaction Symmetries,}
Phys.\ Lett.\ B {\bf 90}, 125 (1980).
%%CITATION = PHLTA,B90,125;%%

%\cite{Dimopoulos:1979es}
\bibitem{Dimopoulos:1979es}
S.~Dimopoulos and L.~Susskind,
\emph{Mass Without Scalars,}
Nucl.\ Phys.\ B {\bf 155}, 237 (1979).
%%CITATION = NUPHA,B155,237;%%

%\cite{Abbott:1981re}
\bibitem{Abbott:1981re}
L.~F.~Abbott and E.~Farhi,
\emph{Are The Weak Interactions Strong?,}
Phys.\ Lett.\ B {\bf 101}, 69 (1981).
%%CITATION = PHLTA,B101,69;%%

%\cite{Abbott:1981yg}
\bibitem{Abbott:1981yg}
L.~F.~Abbott and E.~Farhi,
\emph{A Confining Model Of The Weak Interactions,}
Nucl.\ Phys.\ B {\bf 189}, 547 (1981).
%%CITATION = NUPHA,B189,547;%%


%\cite{Erlich:2005qh}
\bibitem{Erlich:2005qh}
J.~Erlich, E.~Katz, D.~T.~Son and M.~A.~Stephanov,
\emph{QCD and a holographic model of hadrons,}
arXiv:hep-ph/0501128.
%%CITATION = HEP-PH 0501128;%%

%\cite{'tHooft:1973jz}
\bibitem{'tHooft:1973jz}
G.~'t Hooft,
\emph{A Planar Diagram Theory For Strong Interactions,}
Nucl.\ Phys.\ B {\bf 72}, 461 (1974).
%%CITATION = NUPHA,B72,461;%%

%\cite{Harada:2000kb}
\bibitem{Harada:2000kb}
  M.~Harada and K.~Yamawaki,
  \emph{Vector manifestation of the chiral symmetry,}
  Phys.\ Rev.\ Lett.\  {\bf 86}, 757 (2001)
  [arXiv:hep-ph/0010207].
  %%CITATION = HEP-PH 0010207;%%





%\cite{Georgi:1975tz}
\bibitem{Georgi:1975tz}
H.~Georgi and A.~Pais,
\emph{Vacuum Symmetry And The Pseudogoldstone Phenomenon,}
Phys.\ Rev.\ D {\bf 12}, 508 (1975).
%%CITATION = PHRVA,D12,508;%%

%\cite{Coleman:1973jx}
\bibitem{Coleman:1973jx}
  S.~R.~Coleman and E.~Weinberg,
  \emph{Radiative Corrections As The Origin Of Spontaneous Symmetry Breaking,}
  Phys.\ Rev.\ D {\bf 7}, 1888 (1973).
  %%CITATION = PHRVA,D7,1888;%%


%\cite{Holdom:1981rm}
\bibitem{Holdom:1981rm}
B.~Holdom,
\emph{Raising The Sideways Scale,}
Phys.\ Rev.\ D {\bf 24}, 1441 (1981).
%%CITATION = PHRVA,D24,1441;%%

%\cite{Chang:2003zn}
\bibitem{Chang:2003zn}
S.~Chang,
\emph{A `littlest Higgs' model with custodial SU(2) symmetry,}
JHEP {\bf 0312}, 057 (2003)
[arXiv:hep-ph/0306034].
%%CITATION = HEP-PH 0306034;%%

%\cite{'tHooft:1980xb}
\bibitem{'tHooft:1980xb}
G.~'t Hooft, in
\emph{Recent Developments In Gauge Theories:  Nato Advanced Study Institute, Cargese (1979)}, (Plenum, New York, 1980).
%\href{http://www.slac.stanford.edu/spires/find/hep/www?irn=949701}{SPIRES entry}

%\cite{Callan:1984sa}
\bibitem{Callan:1984sa}
C.~G.~Callan and J.~A.~Harvey,
\emph{Anomalies And Fermion Zero Modes On Strings And Domain Walls,}
Nucl.\ Phys.\ B {\bf 250}, 427 (1985).
%%CITATION = NUPHA,B250,427;%%

%\cite{Zumino:1983rz}
\bibitem{Zumino:1983rz}
B.~Zumino, Y.~S.~Wu and A.~Zee,
\emph{Chiral Anomalies, Higher Dimensions, And Differential Geometry,}
Nucl.\ Phys.\ B {\bf 239}, 477 (1984).
%%CITATION = NUPHA,B239,477;%%

%\cite{Manes:1984gk}
\bibitem{Manes:1984gk}
J.~L.~Ma\~{n}es,
\emph{Differential Geometric Construction Of The Gauged Wess-Zumino Action,}
Nucl.\ Phys.\ B {\bf 250}, 369 (1985).
%%CITATION = NUPHA,B250,369;%%

%\cite{Wu:1984pv}
\bibitem{Wu:1984pv}
Y.~S.~Wu,
\emph{Local Cohomology Of Gauge Fields And Explicit Construction Of Wess-Zumino Lagrangians In Nonlinear Sigma Models Over G/H,}
Phys.\ Lett.\ B {\bf 153}, 70 (1985).
%%CITATION = PHLTA,B153,70;%%

%\cite{Fujiwara:1984mp}
\bibitem{Fujiwara:1984mp}
  T.~Fujiwara, T.~Kugo, H.~Terao, S.~Uehara and K.~Yamawaki,
  \emph{Nonabelian Anomaly And Vector Mesons As Dynamical Gauge Bosons Of Hidden
  Local Symmetries,}
  Prog.\ Theor.\ Phys.\  {\bf 73}, 926 (1985).
  %%CITATION = PTPKA,73,926;%%


%\cite{Hill:2004uc}
\bibitem{Hill:2004uc}
C.~T.~Hill and C.~K.~Zachos,
\emph{Dimensional deconstruction and Wess-Zumino-Witten terms,}
Phys.\ Rev.\ D {\bf 71}, 046002 (2005)
[arXiv:hep-th/0411157].
%%CITATION = HEP-TH 0411157;%%


%\cite{Csaki:2003dt}
\bibitem{Csaki:2003dt}
C.~Csaki, C.~Grojean, H.~Murayama, L.~Pilo and J.~Terning,
\emph{Gauge theories on an interval: Unitarity without a Higgs,}
Phys.\ Rev.\ D {\bf 69}, 055006 (2004)
[arXiv:hep-ph/0305237].
%%CITATION = HEP-PH 0305237;%%

%\cite{Chang:2003vs}
\bibitem{Chang:2003vs}
S.~Chang and H.~J.~He,
\emph{Unitarity of little Higgs models signals new physics of UV completion,}
Phys.\ Lett.\ B {\bf 586}, 95 (2004)
[arXiv:hep-ph/0311177].
%%CITATION = HEP-PH 0311177;%%

\end{thebibliography}
\end{document}